\documentclass[12pt,epsfig]{article}                                     
\usepackage{epsfig}          
\usepackage{amssymb}      
\usepackage{amsfonts}          
\newskip\humongous \humongous=0pt plus 1000pt minus 1000pt

\newif\ifdtup

\catcode`\@=11
 
\@addtoreset{equation}{section}
\def\theequation{\thesection.\arabic{equation}}
 
\def\@normalsize{\@setsize\normalsize{15pt}\xiipt\@xiipt
\abovedisplayskip 14pt plus3pt minus3pt%
\belowdisplayskip \abovedisplayskip
\abovedisplayshortskip \z@ plus3pt%
\belowdisplayshortskip 7pt plus3.5pt minus0pt}
 
\def\small{\@setsize\small{13.6pt}\xipt\@xipt
\abovedisplayskip 13pt plus3pt minus3pt%
\belowdisplayskip \abovedisplayskip
\abovedisplayshortskip \z@ plus3pt%
\belowdisplayshortskip 7pt plus3.5pt minus0pt
\def\@listi{\parsep 4.5pt plus 2pt minus 1pt
     \itemsep \parsep
     \topsep 9pt plus 3pt minus 3pt}}
 
\relax

\catcode`@=12
 
\catcode`\@=11
 
\def\section{\@startsection{section}{1}{\z@}{3.5ex plus 1ex minus
   .2ex}{2.3ex plus .2ex}{\large\bf}}
 
\def\thesection{\arabic{section}}    
\def\thesubsection{\arabic{section}.\arabic{subsection}}

\def\appendix{\setcounter{section}{0}
 \def\thesection{Appendix \Alph{section}}
 \def\thesubsection{\Alph{section}.\arabic{subsection}}
 \def\theequation{\Alph{section}.\arabic{equation}}}
 
\def\YGrule{0.4}   
\def\YGbox{6.5}    
\def\SymBoxes#1#2#3#4{\newdimen\un@t \un@t#3%
\raisebox{#1}{\rule{#2\un@t}{#4}\hskip-#2\un@t
\@tempdimb\un@t \advance\@tempdimb by-#4\@tempcntb#2\relax%
\@whilenum{\@tempcntb>0}\do{
\rule{#4}{\un@t}\hskip\@tempdimb \advance\@tempcntb by\m@ne}%
\hskip-#2\un@t \rule[\un@t]{#2\un@t}{#4}%
\rule[\un@t]{#4}{#4}\hskip-#4
\rule{#4}{\un@t}}\hskip-#4}                
\def\Young{\@ifnextchar[{\@Young}{\@Young[0]}}
\def\@Young[#1]#2{\newdimen\YG@unit \YG@unit\YGbox pt%
\newdimen\h@ight \h@ight#1\YG@unit \@tempcnta-1\relax
\@tfor\c@ount:=#2\do{\advance\@tempcnta by\@ne}
\@tempdima\@tempcnta\YG@unit%
\advance\h@ight by\@tempdima\relax     
\@tfor\c@ount:=#2\do{\SymBoxes{\h@ight}{\c@ount}{\YG@unit}{\YGrule pt}%
\@tempdima-\c@ount\YG@unit \hskip\@tempdima%
\advance \h@ight by -\YG@unit}         
\@tempdima\YG@unit \multiply\@tempdima by\@car#2\@nil %
\hskip\@tempdima}                      
\def\YoungTab{\@ifnextchar[{\@YoungIdx}{\@YoungIdx[0]}}
\def\@YoungIdx[#1]{\@ifnextchar[{\@iYoungIdx[#1]}{\@iYoungIdx[#1][\@empty]}}
\def\@iYoungIdx[#1][#2]#3{%
\newdimen\YG@unit \YG@unit\YGbox pt\newdimen\YG@rule \YG@rule \YGrule pt
\newcount\c@ount \c@ount\z@ \newdimen\skip@wd \unitlength\@ne pt
\newdimen\h@ight \h@ight#1\YG@unit \@tempcnta\m@ne\relax
\@tfor\d@um:=#3\do{\advance\@tempcnta by\@ne}
\@tempdima\@tempcnta\YG@unit%
\advance\h@ight by\@tempdima\relax
\@tfor\@idxlist:=#3\do{
\@tempcnta\z@\hskip.5\YG@rule\relax 
\@for\@idx:=\@idxlist\do{
\raisebox{\h@ight}{\makebox(\YGbox,\YGbox){#2$\@idx$}}
\advance\@tempcnta by\@ne}\hskip-.5\YG@rule%
\@tempdima-\@tempcnta\YG@unit \hskip\@tempdima%
\ifnum\c@ount=\z@ \skip@wd-\@tempdima\fi \relax
\SymBoxes{\h@ight}{\@tempcnta}{\YG@unit}{\YG@rule}%
\hskip\@tempdima \advance\h@ight by -\YG@unit
\advance\c@ount by\@ne}
\hskip\skip@wd}                      

\begin{document}

\newcommand{\beq}{\begin{equation}}
\newcommand{\eeq}{\end{equation}}
\newcommand{\bea}{\begin{eqnarray}}
\newcommand{\eea}{\end{eqnarray}}
\newcommand{\beas}{\begin{eqnarray*}}
\newcommand{\eeas}{\end{eqnarray*}}
\newcommand{\defi}{\stackrel{\rm def}{=}}
\newcommand{\non}{\nonumber}
\newcommand{\bquo}{\begin{quote}}
\newcommand{\enqu}{\end{quote}}
\def\de{\partial}
\def\Tr{ \hbox{\rm Tr}}
\def\const{\hbox {\rm const.}}
\def\o{\over}
\def\im{\hbox{\rm Im}}
\def\re{\hbox{\rm Re}}
\def\bra{\langle}\def\ket{\rangle}
\def\Arg{\hbox {\rm Arg}}
\def\Re{\hbox {\rm Re}}
\def\Im{\hbox {\rm Im}}
\def\diag{\hbox{\rm diag}}
\def\longvert{{\rule[-2mm]{0.1mm}{7mm}}\,}
\def\Z{\mathbb Z}
\begin{titlepage}
{\hfill     ULB-TH-04/11, IFUP-TH/2004-5 } 
\bigskip 
\bigskip

\begin{center}
{\large  {\bf  
NONABELIAN 
  MONOPOLES  
 } } 
\end{center}

\renewcommand{\thefootnote}{\fnsymbol{footnote}}
\bigskip
\begin{center}
{\large  Roberto AUZZI $^{(1,3)}$ , Stefano BOLOGNESI $^{(1,3)}$, \\
 Jarah EVSLIN $^{(4)}$, Kenichi KONISHI $^{(2,3)}$,     Hitoshi
 MURAYAMA $^{(5)}$\footnote{On leave of absence from \it Department of
 Physics, University of California, Berkeley, CA 94720, USA}
 \vskip 0.10cm
 }
\end{center}

\begin{center}
{\it      \footnotesize
Scuola Normale Superiore - Pisa $^{(1)}$,
 Piazza dei Cavalieri 7, Pisa, Italy \\
Dipartimento di Fisica ``E. Fermi" -- Universit\`a di Pisa $^{(2)}$, \\
Istituto Nazionale di Fisica Nucleare -- Sezione di Pisa $^{(3)}$, \\
Via Buonarroti, 2, Ed. C, 56127 Pisa,  Italy,   \\  
 
International Solvay Institutes, 
Universit\'e Libre de Bruxelles,  \\
Campus Plaine C.P. 231, B-1050 Bruxelles, Belgium$^{(4)}$  \\ 
     Institute for Advanced Study, Princeton, NJ 08540,     USA   $^{(5)}$
}    

\end {center}

\setcounter{footnote}{0}  

\noindent  
{\bf Abstract:}  
  
{ We study   topological as well as dynamical properties of  BPS nonabelian magnetic  
  monopoles of Goddard-Nuyts-Olive-Weinberg type in $ G=SU(N)$,
  $USp(2N)$ and $SO(N)$ gauge theories, spontaneously broken to
  nonabelian subgroups $H$.  We find that monopoles transform under
  the group dual to $H$ in a tensor representation of rank determined
  by the corresponding element in $\pi_1(H)$.  When the system is
  embedded in a ${\cal N}=2$ supersymmetric theory with an appropriate
set of flavors with appropriate bare masses,    the BPS monopoles constructed semiclassically
  persist in the full quantum theory.  This result supports the
  identification of ``dual quarks'' found  at $r$-vacua  of ${\cal N}=2$  theories  with the nonabelian
  magnetic monopoles.  We present several consistency checks of our monopole spectra. }

\vfill  
 
\begin{flushright}
\today
\end{flushright}
\end{titlepage}

\bigskip

\hfill{}

\section{Introduction}

After many years of investigations \cite{Lb}-\cite{BS}  the dynamical
roles that nonabelian monopoles can play in physically interesting
gauge theories are finally being understood.  Although their presence
in classical examples of conformally invariant ${\cal N}=4$ theories
and in more recent ${\cal N}=1$ supersymmetric models with Seiberg
duals \cite{Sei} is by now well known, monopoles with nonabelian gauge
symmetry manifest themselves most clearly (e.g., as the infrared
degrees of freedom) in softly broken ${\cal N}=2 $ gauge theories
coupled to fundamental matter hypermultiplets
\cite{SW1}-\cite{curves}.

In particular, a series of papers on this class of theories with gauge
groups $SU(N)$, $USp(2N)$ and $SO(N)$ and various numbers of flavors
of matter multiplets \cite{APS}-\cite{CKKM} systematically
investigated the infrared fate of these monopoles in every vacuum.
The ``dual quarks" appearing as the low-energy degrees of freedom in
certain vacua of these theories, which are charged under the unbroken
nonabelian $SU(r) \subset G$, can be identified \cite{BK} with the
semiclassical monopoles studied earlier by Goddard, Nuyts and Olive
\cite{GNO} and also by E. Weinberg \cite{EW}.  {\it All } of the
confining vacua of strongly coupled $USp(2N)$ and $SO(N)$ theories
with flavors and without bare quark masses involve these objects in a
deformed SCFT \cite{CKM,CKKM,AGK}.  

Recently, with A. Yung, we have explored the properties of
nonabelian BPS {\it vortices } appearing in the same class of models
\cite{ABEKY}.  In particular, continuous transformations among the
degenerate vortex solutions (vortex zero modes) were explicitly
constructed, showing the true nonabelian nature of these solitons.
The analysis is done in a region of large bare quark masses where the
semi-classical approximation holds, and yet the whole result is
quantum mechanically valid when the model is embedded in the ${\cal
  N}=2$ theory \cite{ABEKY}.
\footnote{There are papers related to ours by Hanany and Tong \cite{HT}, and also  by Kneipp\cite{Kneipp}. 
Interesting recent articles  \cite{Tong, ShifYun} furthermore    relate the vortex dynamics 
to that of 4D  gauge theory itself.   }

In the present paper, our attention will be focused on the topological
and dynamical properties of nonabelian monopoles themselves, in a wide
class of ${\cal N}=2$ gauge theories with $SU(N)$, $USp(2 N)$ and
$SO(N)$ gauge groups, spontaneously broken to various nonabelian
subgroups $H$.  In particular for each symmetry breaking pattern the
minimal monopoles are identified and their nonabelian and abelian
charges determined.  The monopole-vortex flux matching argument given
in \cite{monovort} is crucial in demonstrating, albeit indirectly,
that a continuous set of these monopoles exist, forming a multiplet of
the dual gauge group, ${\tilde H}$.

This paper is organized as follows. We start our analysis
(Section~\ref{NAMonopoles}) by reviewing the properties of BPS
``nonabelian" monopole solutions in a bosonic theory with a
spontaneously broken $SU(N+1) \to SU(N) \times U(1)$ symmetry.  We
then show that when the model is embedded in a ${\cal N}=2$
supersymmetric theory \cite{SW1}-\cite{curves}, the semiclassical
solutions of the bosonic model acquire a quantum mechanical meaning.
The quantum mechanical aspects of nonabelian monopoles and in
particular the importance of the massless flavors in the underlying
theory, are discussed more thoroughly and in a more general context   in Section~\ref{sec:quantum}.
In Section~\ref{sec:homotopy} we establish their topological stability
and determine their abelian magnetic charges with respect to the
minimal Dirac quantum, based on homotopy group analysis,  for $G= SU(N),$ $USp(2N)$ or $SO(N)$ and for various choices of $H$. 
   In
Section~\ref{sec:Explicit} we present an explicit construction of these minimal monopoles. 
We summarize our results    in  Section~\ref{sec:conclusion},   presenting also some  crosschecks of  consistency of our
results.  In \ref{sec:General}, the general formulae due to E.
Weinberg and  to Goddard-Nuyts-Olive     are reviewed   and somewhat
streamlined, which helps as a reference for other parts of the    paper.
For completeness and  for convenience   we review also   the root vector systems and Cartan
subalgebras of $SU$, $SO$ and $USp$ Lie algebras in \ref{sec:Roots}.

\section{Nonabelian BPS   Monopoles in $SU(N + 1 )$   Theories \label{NAMonopoles}}

For illustration we begin our discussion by briefly reviewing the  properties of  the monopoles arising in a
system with symmetry breaking $SU(N+1) \to SU(N) \times U(1)$. 

\subsection {Bosonic   $SU(N + 1 )$ theory with an adjoint scalar \label{sec:ordinary}}  

We are interested in the standard  $SU(N + 1 )$    model
\begin{equation}  {\cal L}=  { 1\o 4 g^2}  (F_{\mu \nu}^A)^2  +  { 1\o g^2}  |({\cal D}_{\mu} \phi)^A|^2 - V(\phi),   \end{equation}
with a complex adjoint scalar field $\phi$.  Let us assume that the potential is minimized by an adjoint scalar VEV of the form
\begin{equation}    \bra \phi  \ket  =
 \pmatrix{  v     &  0  &  \ldots &    0
  \cr  0   & v    & \ldots   & 0
 \cr  \vdots  &\vdots& \ddots & \vdots
\cr  0    & 0   &  \ldots  & -N v     } =
\pmatrix{  v \cdot {\bf 1}_{N\times N}     &
  \cr    & -N v     }.   
\label{phivev}  \end{equation}
Such a  VEV breaks the gauge symmetry as
\begin{equation}   SU(N+1)  \to   {SU(N)  \times U(1)  \o {\mathbb Z}_N }=U(N),
\label{Symbr}\end{equation}
where the $ {\mathbb Z}_N $  factor arises because the $n$th roots of unity in $U(1)$ also lie in $SU(N)$.

The stability of a monopole is guaranteed by the topological
nontriviality of both its $SU(N+1)$-valued adjoint scalar Higgs field
and its gauge fields.  The Higgs VEV, evaluated on a 2-sphere
surrounding any monopole, provides a map from the 2-sphere to the
space of orbits of the unbroken gauge group $U(N)$ in the original
$SU(N+1)$.  The Higgs VEV is therefore a representative of
$\pi_2(SU(N+1)/U(N))$.
 
On the other hand, The gauge field configuration describes a $U(N)$
gauge bundle over the spacetime with the monopole deleted, which can
be contracted to a 2-sphere.  Such a gauge bundle may be trivialized
over the northern and southern hemispheres, and so is entirely
characterized by the transition function on the equator.  The
transition function is a map from the equatorial circle to the gauge
group $U(N)$, and so it is a representative of $\pi_1(U(N))$.

The long exact sequence for homotopy groups of fibrations assures that
these two classifications of monopoles agree.  The two homotopy groups
are both the group of integers
\begin{equation} 
  \pi_2\left({SU(N+1)\o  U(N)}\right)  = \pi_1(U(N)) = {\mathbb Z} 
\label{homotopy}  
\end{equation}  
and so these monopoles are topologically stable and may carry any
integral charge.  We will see later that if the original gauge
symmetry $G$ is not simply connected then the long exact sequence
yields more topologically distinct gauge field configurations than
Higgs configurations.  These extra gauge field configurations are
singular Dirac-like monopoles that exist in the $G$ gauge theory even
without symmetry breaking, and are not the nonsingular monopoles of
interest in this paper.  For semi-simple $G$ the fundamental group is
torsion and so the extra monopole charges are pure torsion and the
extra monopoles are not BPS.

The mass of a BPS monopole may be read from the Hamiltonian
\begin{equation}  
  H = \int d^3x  \, \left[   { 1\o 4 g^2}  (F_{i j}^A)^2  +  { 1\o g^2}
  |{\cal D}_{i} \phi^A|^2  \right] = \int d^3x \, \left[   { 1\o 4 g^2}
  (F_{i j}^A)^2  +  { 1\o 2   g^2}  |{\cal D}_{i} \phi^A|^2  \right]  
\end{equation}
where in the final expression we have kept only the real part of
$\phi^A$.  Rewriting the Hamiltonian as
\begin{equation}
  H=     \int d^3x  \, \left[   { 1\o 4 g^2} \left|  F_{i j}^A  \pm
  \epsilon_{ijk}     ({\cal D}_{k} \phi)^A  \right|^2  \pm { 1\o 2}
  \de_k  (\epsilon_{ijk}   F_{i j}^A  \phi^A)\right] \label{BPSHamil}
\end{equation}
BPS monopole \cite{PS} configurations are seen to satisfy the {\it
  nonabelian } Bogomolny equations:
\begin{equation}
  B_k^A= -   ({\cal D}_{k} \phi)^A; \qquad  B_k^A = {1 \o 2}\,
  \epsilon_{ijk}  F_{i j}^A \label{nbe}.
\label{BPSeqs}
\end{equation}
The BPS bound on the monopole mass is given by the following integral,
performed on the 2-sphere at $r=\infty$:
\begin{equation}  
  H=  \int dS \cdot  (\phi^A  {\bf B}^A ) = { 2 \pi \o g} \,3  \, v \,
  k, \qquad k =1,2, \ldots. \label{bb} 
\end{equation}

More explicitly,   a   monopole solution can be  found by
choosing   an $SU(2)$ subgroup:
{\small\begin{equation}
S_1=\frac{1}{2} \pmatrix{  0     &  0  &  \ldots &    1
  \cr  0   & 0    & \ldots   & 0
 \cr  \vdots  &\vdots& \ddots & \vdots
\cr  1    & 0   &  \ldots  & 0     };\quad
 S_2=\frac{1}{2} \pmatrix{  0     &  0  &  \ldots &    -i
  \cr  0   & 0    & \ldots   & 0
 \cr  \vdots  &\vdots& \ddots & \vdots
\cr  i    & 0   &  \ldots  & 0     }; \quad 
     S_3=\frac{1}{2} \pmatrix{  1     &  0  &  \ldots &
   0
  \cr  0   & 0    & \ldots   & 0
 \cr  \vdots  &\vdots& \ddots & \vdots
\cr  0    & 0   &  \ldots  & -1     }.  \label{first}   
\end{equation}}
The monopole  solution
is then given by  \cite{EW,BK}:
{\small   \begin{equation}
\phi=\pmatrix{  -\frac{N-1}{2} v     &  0  &  \ldots &    0
 &0 
  \cr  0   & v    & 0 & \ldots    &0
\cr 0 & 0 & v & \ldots & 0
 \cr  \vdots  &\vdots& \vdots& \ddots & \vdots
\cr  0    & 0  &0 &  \ldots  & -\frac{N-1}{2} v   }
+(N+1)v  (\vec{S} \cdot \widehat{r}) \phi(r),
\end{equation} }
\begin{equation}
\vec{A}(r)=\vec{S} \wedge \widehat{r}  \, \frac{A(r)}{g}
\end{equation}
where $\phi(r)$ and $A(r)$ are 't Hooft-BPS functions with
$\phi(\infty)=1$, $\phi(0)=0$, $A(\infty)=-1/r$.  To compute the mass,
using Eq.~(\ref{bb}) one needs the following property:
\begin{equation}
  { 1\o 2}   \epsilon^{ijk} F_{ij} \widehat{r}_k=
  -\frac{\vec{S}\cdot \widehat{r}}{g \,  r^2}, 
\end{equation}
yielding  the result
\begin{equation}
  M=\frac{2 \pi (N+1) \, v}  {g}. \label{sunmass}
\end{equation} 

In order to calculate the abelian charge with respect to the $U(1)$
factor in Eq.(\ref{Symbr}) we need to first calculate the total
magnetic flux sourced by the minimal monopole.  Normalizing the flux
by the norm of $\phi$, one finds
\begin{equation}   
  F_m=  \int_{S^2}   d{\bf S}    \cdot {\Tr  \,  \phi
    \,  {\bf  B}  \o { 1\o \sqrt {2} }  ( \Tr \phi^2)^{1/2} }    = { 2 \pi
    (N+1)  \o   {g\sqrt 
      {N(N+1)} /\sqrt{2} }}
  =  {{2 \pi}\o {g}} \,
  \sqrt{2(N+1)  
    \o N}.
\label{Mflux}
\end{equation}
This should be equal to $4 \pi \, g_m$, so
\begin{equation}   g_m=   \frac{1}{g} \sqrt{N+1 \o  2 \, N} .
\end{equation}
On the other hand, the electric coupling of the $A^0_{\mu}$ field to
the matter in the fundamental representation of $SU(N+1)$ is through
the minimum coupling constant
\begin{equation}  e_0=   { g  \o  \sqrt{2N(N+1)}},
\end{equation}
as
\begin{equation}   t^0  =  { 1 \o  \sqrt{2N(N+1)}}
\pmatrix{  {\bf 1}_{N\times N}     &
  \cr    & -N      }.
\end{equation}
Thus the minimum magnetic charge, in terms of the unit electric charge, is
\begin{equation}  g_m=   { 1 \o 2\,N\,  e_0}
\label{suncharge}\end{equation}
which is $1/N$ of the charge of Dirac's $U(1)$ monopole \cite{Dirac}.
In Sec.\ref{sec:homotopy} we will see that this factor of $N$ is the
degree of the embedding of the fundamental group of the unbroken
$U(1)$ into that of the unbroken gauge group.

Clearly, the choice made above (\ref{first}) is nothing but one of the
$N$ possibilities.  By using the $SU(2)$ subgroups lying in the $(i,
N+1)$ - $2 \times 2$ subspaces, $i=1,2,\ldots, N$, we finds $N$
degenerate monopoles with identical masses and charges.  This is the
right multiplicity for these monopoles to belong to the fundamental
representation of the dual $SU(N)$ magnetic group.

\subsection{Embedding the System  in ${\cal N}=2$  }

The fact that the  monopoles associated with the symmetry breaking 
$   G   \,\,\,{\stackrel {\bra \phi \ket    \ne 0} {\longrightarrow}}     \,\,\, H  $
semiclassically form a  degenerate  multiplet, however,    does not in itself prove that they are nonabelian monopoles.
The main problem   is that  the  ``unbroken" gauge group $H$  (in the case just considered,  $U(N)$)   can dynamically break down to an  abelian subgroup.
Whether such a dynamical breaking occurs depends on the details of the quantum system and it is in general difficult to decide  what actually takes
place. 

In ${\cal N}=2$   supersymmetric  models  a definite answer can be given.     If the model above  is embedded in the pure  ${\cal N} =2$   theory 
\begin{equation}
{\cal L}=     {1\over 8 \pi} \im \, {\tau}_{cl} \left[\int d^4 \theta \,
\Phi^{\dagger} e^V \Phi +\int d^2 \theta\,{1\o 2} W W\right], 
\label{lagrangian}
\end{equation}
($
{\tau}_{cl} \equiv  {\theta_0 \o \pi} + {8 \pi i \o g_0^2}
 $),  the $SU(N)$   sector left ``unbroken" by the   adjoint scalar VEV (\ref{phivev})  describes a pure  ${\cal N} =2$  $SU(N)$  theory, 
which becomes strongly coupled at low energies and is dynamically broken to the maximal Abelian subgroup $U(1)^{N-1}$ \cite{SW2,curves,DS}.

In order to preserve an unbroken subgroup $H$  we couple the theory to  $N_f$
  hypermultiplets  (quarks).    
The Lagrangian of this theory has the structure
\begin{equation}
{\cal L}=     {1\over 8 \pi} \im \, S_{cl} \left[\int d^4 \theta \,
\Phi^{\dagger} e^V \Phi +\int d^2 \theta\,{1\o 2} W W\right]
+ {\cal L}^{(quarks)}  +  \int \, d^2 \theta \,\mu  \,\Tr  \Phi^2;  
\label{lagrangianbis}     
\end{equation}
\begin{equation} {\cal L}^{(quarks)}= \sum_i \, [ \int d^4 \theta \, \{ Q_i^{\dagger} e^V
Q_i + {\tilde Q_i}  e^{-V} {\tilde Q}_i^{\dagger} \} +  \int d^2 \theta
\, \{ \sqrt{2} {\tilde Q}_i \Phi Q^i    +      m   {\tilde Q}_i Q^i   \}
\label{lagquark}
\end{equation}
where $m$ is the (common)  bare mass of the quarks and we have defined the complex coupling constant
\begin{equation}
S_{cl} \equiv  {\theta_0 \o \pi} + {8 \pi i \o g_0^2}.  
\label{struc}
\end{equation}   
The parameter $\mu$ is the mass of the adjoint chiral multiplet, which breaks the supersymmetry to ${\cal N}=1$.

In order to discuss unconfined monopoles, however,     one   must set $\mu=0$ and so preserve the full ${\cal N}=2$
supersymmetry. 
     After eliminating the auxiliary fields  the bosonic Lagrangian becomes 
\begin{equation}  
  {\cal L}=  { 1\o 4 g^2}  F_{\mu \nu}^2  +  { 1\o g^2}  |{\cal
  D}_{\mu} \phi|^2 +   \left|{\cal D}_{\mu}
  Q\right|^2 + \left|{\cal D}_{\mu} \bar{\tilde{Q}}\right|^2 +  {\cal
  L}_1+  {\cal L}_2,  
\label{Lag}\end{equation}
where
\begin{eqnarray}    
  {\cal L}_1 &=&  -   { 1\o 8 }  \, \sum_A  \left[ { 1\o g^2 }  (-  i)
  f_{ABC} \, \phi^{\dagger}_B  \phi_C +  Q^{\dagger} t^A  Q -  {\tilde
    Q} t^A   {\tilde Q}^{\dagger} \right]^2    \non \\
  &=&     -   { 1\o 8 }  \, \sum_A  \left( t^A_{ij} \,  \left[ { 1\o g^2 }
  (-2)  \, [\phi^{\dagger},   \phi]_{ji}  +  Q^{\dagger}_j   Q_i -
  {\tilde Q}_j   {\tilde Q}^{\dagger}_i \right]\right)^2; 
\end{eqnarray}
\begin{eqnarray}    
  {\cal L}_2 &=&   - g^2 | 
  \sqrt 2   \, {\tilde Q} \, t^A  Q |^2   -   {\tilde Q } \,   [ m
  + \sqrt2  \phi   ] \,  [ m    + \sqrt2  \phi   ]^{\dagger}  \,
  {\tilde Q}^{\dagger}  \non \\  
  &-& Q^{\dagger} \, [ m    + \sqrt2  \phi   ]^{\dagger}    \, [ m
  + \sqrt2  \phi   ]  \, Q.   
\end{eqnarray}
In the construction of the monopole solutions  we shall consider only the VEVs and fluctuations around them which satisfy 
\begin{equation}   [\phi^{\dagger},   \phi]=0, \qquad    Q_i =  {\tilde Q}^{\dagger}_i,
\end{equation}
and hence   ${\cal L}_1$  can be set to zero. 

We are interested in the vacua which would survive the above ($\mu\neq
0$) perturbation to ${\cal N}=1$.  These vacua are parametrized by the
integer $r$, which is the rank of the unbroken nonabelian gauge
symmetry plus one \cite{APS,CKM}.  In order to exhibit the symmetry
breaking $SU(N+1) \to SU(N) \times U(1)$ one may choose the adjoint
VEV to be
\begin{equation}   \bra \phi  \ket =- { m \o \sqrt {2} }  \, \diag \,(
1,1, \ldots, 1, - N), \label{adjvevbis} 
\end{equation}
namely, as in Eq.(\ref{phivev}) with $v= - { m \o \sqrt {2} }$.
Together with the vanishing squark VEVs
\begin{equation}  \bra Q_i \ket =   \bra  {\tilde Q}_i \ket = 0,  
\label{squark}\end{equation}
this is easily seen to give vanishing contribution to the bosonic
Lagrangian (\ref{Lag}).  (\ref{adjvevbis}) and (\ref{squark})
represent a supersymmetric vacuum as ${\cal L}_1={\cal L}_2=0.$

One is left with the task of minimizing the first two terms of the
bosonic Lagrangtian (\ref{Lag}): the rest of the discussion is exactly
as in Sec.~\ref{sec:ordinary}: semiclassically nothing changes.
Quantum mechanically, however, there is an important difference.  In
the vacuum characterized by the VEV (\ref{adjvevbis}) there are $N_f$
massless quarks (and squarks) in the fundamental representation of
$SU(N)$.  This ensures that the subgroup $SU(N)$ is non-asymptotically
free and remains unbroken in the infrared  if $2(N+1) > N_f \ge 2N.$ 
   This strong restriction on the number
of flavors may be significantly relaxed in cases in which the unbroken
group is smaller.  In this way one finds that the only real
restriction is that the number of flavors be at least equal to $2r $
if the monopole transforms in the fundamental representation of
$SU(r)$.  (See {\it e.g.,} Eq.(\ref{betadual}).)

\section{ Quantum   Nonabelian  Monopoles \label{sec:quantum}}

The above example of the $SU(N+1)$ model nicely illustrates the fact
that a semiclassical treatment alone is not enough to ensure that the
set of apparently degenerate monopoles associated with the symmetry
breaking $ G \,\,\,{\stackrel {\bra \phi \ket \ne 0}
  {\longrightarrow}} \,\,\, H $ are truly nonabelian.  The reason
is that the ``unbroken" gauge group $H$ may well dynamically break
down to an abelian subgroup.  If this occurs, one has only an {\it
  approximately} degenerate set of monopoles whose masses differ by
e.g., $O({\Lambda^2 \o \bra \phi \ket})$.  For this reason, the very  concept of 
nonabelian monopoles  is never really semi-classical, in sharp
contrast to the case of abelian monopoles.
Only if the ``unbroken" gauge group $H$ is not further broken
dynamically do the unconfined (topologically stable) nonabelian
monopoles and dual gauge bosons appear in the quantum theory.

Another subtlety is that it is not justified to study the system $ G
\,\,\,{\stackrel {\bra \phi \ket \ne 0} {\longrightarrow}} \,\,\, H $
with a nonabelian subgroup $H$ as a limiting situation of a maximal
breaking,  $ G \,\,\,{\stackrel {\bra \phi \ket \ne 0}
  {\longrightarrow}} \,\,\,   U(1)^R,$ where $R$ is the rank of the
group $G$, by letting some of the eigenvalues of $\bra \phi \ket $ to
coincide, as is sometimes done in the literature.  To do so would
introduce fictitious degrees of freedom corresponding to massless,
infinitely extended ``solitons".  In this limit all fields tend to
constant values and so in fact these are not solitons at
all.\footnote{This is analogous to what would happen to the 't Hooft -
  Polyakov monopole of the spontaneously broken $SU(2){\stackrel {v}
    {\longrightarrow}} U(1) $ theory, if one were to apply the
  semi-classical formulae na\"{i}vely in the limit $v\to 0$.} Indeed,
in the case $G=SU(N)$, such ``massless monopoles" do not represent any
topological invariant as the fundamental group of any restored $SU(N)$
is trivial.

It is hardly possible to overemphasize the importance of the fact
\cite{GNO,EW,BK} that nonabelian monopoles, if they exist quantum
mechanically, transform as irreducible multiplets of the dual group
${\tilde H}$, {\it not } under $H$ itself.  Monopoles
transforming under the dual group evade the ``no-go" theorem of \cite{CDyons} which
is a topological obstruction to the existence of monopoles
transforming under the original group.  The distinction between
monopoles transforming under the original versus the dual group is
particularly evident in the cases of $USp(2N)$ or $SO(2N+1)$ gauge
theories which will be considered below.  For instance, in the system
with spontaneous symmetry breaking, $USp(2N+2) \rightarrow USp(2N)
\times U(1)$, we find (see Section~\ref{sec:Explicit},
\ref{sec:General}) that the semi-classical (hence candidate)
nonabelian monopoles form a degenerate $(2N+1)$-plet.  While there are
no $(2N+1)$-dimensional representations of $H=USp(2N) $, the
fundamental representation of the dual group ${\tilde H} = SO(2 N +1)$
has precisely the desired dimension.  Analogously, in the system with
gauge symmetry breaking $SO(2N+3) \rightarrow SO(2 N+1) \times U(1)$
we find $2N$ degenerate monopoles.  Again this is the right
multiplicity for ${\tilde H} = USp(2N)$, the group dual to $SO(2N+1)$.

The theorem of \cite{CDyons} however does have an implication.  It
tells us that the gauge symmetry of the system in the presence of
nonabelian monopoles is not a direct product  
$$     H \otimes   {\tilde H}  
$$
as sometimes suggested, but rather  $ {\tilde H} $  (or
something else, if  the physical degrees of freedom relevant
is a dyon).\footnote{A similar conclusion is reached by Bais and Schroers \cite{BS} who  did a careful analysis of 
surviving generators    for $SU(N)$  theories.}   
   The doubling of the
gauge symmetry does not take place.

{\it  The strongest evidence }        so far that nonabelian monopoles do exist
quantum mechanically and can even become dominant degrees of freedom
in the infrared, comes from the ${\cal N}=2$ theories of
Refs.~\cite{APS}-\cite{BK}.  For instance, in the so-called $r$-vacua
($r < n_f/2$) with an effective $SU(r) \times U(1)^{N-r+1} $ gauge
group in ${\cal N}=2$, $SU(N+1)$ gauge theories, a set of magnetic
particles, some in the fundamental representation of $SU(r)$, and some
others singlets of $SU(r)$, appear as light, low-energy effective
degrees of freedom.   Their charge structure is identical to those
  found for the BPS semiclassical monopoles of minimal mass
  \cite{BK}.\footnote{In fact, the monopoles can acquire flavor
  quantum numbers due to fermion zero modes in anti-symmetric
  representations of $SU(N_f)$ \cite{JR,BK},   and the ``dual quarks'' in the effective
  Lagrangian \cite{APS} at those points on the moduli space can
  naturally be in the fundamental representation of $SU(N_f)$.  This
  makes the identification of ``dual quarks'' at these ``$r$-vacua'' with
  non-abelian monopoles possible.  Why this particular   flavor representation
  remains,   however, is a dynamical question. }  When an adjoint mass
perturbation breaking supersymmetry from ${\cal N}=2 $ to ${\cal N}=1$
is added, these nonabelian monopoles condense and give rise to dual
superconductors (confinement) of nonabelian type, together with flavor
symmetry breaking.

Identical sets of nonabelian monopoles appear in the $r$- vacua of
${\cal N}=2$ $SO(N)$ and $USp(2N)$ theories \cite{CKM,CKKM},  with nonzero bare quark
masses.\footnote{As noted in \cite{Eguchi} and in \cite{CKM}, this fact   is
  related to the universality of these SCFTs appearing at the related
  singularities of the ${\cal N}=2$ theories \cite{SCFT}.}  Again, a
detailed quantum analysis shows the presence (and the crucial role) of
monopoles appearing from the breaking
\begin{equation}   SO(2N),    \,\, SO(2N+1)  \,\,  {\hbox {\rm or}} \,\, USp(2N)  \to     SU(r)  \times   U(1)^{N-r+1}, \qquad  r < { n_f \o 2}
\label{scftIR}\end{equation}
with the same charges as the semiclassical
Goddard-Olive-Nuyts-Weinberg monopoles, but massless and playing the
role of the order parameters of confinement.

These examples   illustrate very clearly the crucial role played by
the massless flavors in the quantum theory.  Without massless flavors
the systems go through dynamical (gauge) symmetry breaking and the
nonabelian gauge symmetry of the monopoles is destroyed.

In the presence of a sufficient number of flavors the dual $SU(r)$
gauge symmetry in (\ref{scftIR}) is exact in the infrared, thanks to
the renormalization effects.  Due to the flavor charge of the
nonabelian monopoles the dual $SU(r)$
coupling constant is infrared-free
\beq      b_0^{(dual)} \propto   -  2 \, r  +   n_f  >  0,   
\label{betadual}   \eeq
while  the original electric theory  is asymptotically free: 
\beq    b_{0} \propto  -  2 \, N +    n_f  <     0, \qquad    -   \, N +2 +    n_f  <     0, \qquad -  2 \, N -2  +    n_f  <     0, \qquad  
\label{betafund}   \eeq
   (for $SU(N),$  $SO(N),$ and $USp(2N)$, respectively).   This is how we understand the actual occurrence of the quantum  $r$ - vacua in the
${\cal N}=2,$ $SU(N),$  $SO(N)$ and $USp(2N)$  gauge theories,   only  in systems   with flavors and with the value $r$ limited by   $\left[ {n_f \o
2}\right]$.  When such a sign flip is not possible for some reason,  such as in pure ${\cal N}=2$ SYM or in  generic vacua of ${\cal N}=2$ 
theories,       dynamical abelianization is expected to   and   indeed does    take   place.    
 

We are  thus   led to draw   the following conclusion.    
For systems with a 
symmetry breaking $G \to H$,  the  general   criterion for the persistence of nonabelian monopoles in the quantum
theory  is that    the system   be such that  it produces, upon  symmetry breaking,   a sufficient number of  massless flavors carrying  charges in
$H$,  so as to protect  the latter group from becoming strongly-coupled in the infrared and  from dynamically breaking itself. 

Let us illustrate this for different cases considered in this paper.   
A detailed and concrete discussion was given  in Section \ref{NAMonopoles}  ~  for  the system   $SU(N+1) \to  SU(N) \times U(1)$, which easily
generalizes  to a more general  breaking patterns    $SU(N) \to  SU(r) \times U(1)^{N-r+1}.$

The cases   $USp(2N) \to  SU(r) \times U(1)^{N-r+1}$  is  dealt with   also quite straightforwardly (see Sec. (3.2) of Ref.~ \cite{CKM}).   By  embedding the
system in the
${\cal N}=2$  context,  all the $n_f$  hypermultiplets can  be given  an equal  nonzero bare mass, and   the breaking is achieved by the adjoint scalar VEV
of the form, 
  \beq
\phi = {1\o \sqrt2}  \, \diag \, (im_1, im_2, \ldots , im_r,0,\ldots,
-im_1, -im_2,
\ldots -im_r,0,\ldots ) \, ;
\eeq
where $m_i \to m$.  The condition  $ r >  {n_f \o 2}$   guarantees  that the subgroup  $SU(r)$  survives in the infrared;   this last condition is
clearly  compatible with the asymptotic freedom of the original    $USp(2N)$   theory  ($n_f  <  2N  +2$).\footnote{Of course, one  could drop  this requirement and work with a larger
number of flavors, so that the unbroken
group $H$ is always non aymptotically free,  if  one is satisfied with a semiclassical analysis ($m\gg  \Lambda$).  The point is that   interesting
things - such as light nonabelian monopoles - happen when the underlying theory becomes strongly coupled in the infrared, and to study such a system
one must tune  the bare masses
$m$  to
$O(\Lambda)$  or even to zero.}

In order to have    the breaking   $USp(2N+2) \to  USp(2N) \times U(1)$  (or  a smaller   $USp$ factor), a different setting is needed.  We give only  one of
the hypermultiplets a large bare mass $m$,  and cancel   it by  
  \beq
\phi = {1\o \sqrt2}  \, \diag \, (0,\ldots,0,  i m; 
0,\ldots,0,  -i m) \, ;
\eeq
so as to satisfy the vacuum equations.   The unbroken   $USp(2N)$  survives in the infrared if    $\, n_f - 1   \ge  2N+2 \, $  while the original  $USp(2N+2)
$ is asymptotically free   for   $n_f  <   2N+4. $  So in this case the only possible value is  $n_f=    2N+3$;    for a smaller  $USp$ factor   the
condition is less severe.

In the case of       the  symmetry breaking,
$SO(2N+2) \to  SO(2N) \times U(1)$,   we again   embed the  system in a ${\cal N}=2 $  theory with  $n_f$ hypermultiplets of which one has a large 
bare mass $m$, while others have none.   When  the adjoint scalar VEV is of an appropriate form   (Eq.(\ref{vevson}) below,  with  $v= i m/\sqrt {2}$),    the
$SO(2N)$ components of the other    $n_f -1$  hypermultiplets  remain   massless.  Thus for  $n_f -1 \ge   2N-2$  (which is compatible  with
the requirement that the original theory is asymptotically free, i.e.,   $n_f < 2 N $)  the unbroken   group $SO(2N)$  is non asymptotically free.  
For more general patterns of breaking,  $SO(2N+2) \to  SO(2 \, r) \times U(1)^{N-r+1} $,  it is even    easier    to arrange the ${\cal N}=2$ system
so that  the unbroken $ SO(2 \, r) $  is non-asymptotically free (Eq.~(\ref{betadual})  {\it vs} Eq.~(\ref{betafund}) ).

In the case of the diagonal breaking    
$SO(2N) \to  U(N)$ considered also below,   we find that the standard  embedding the system in   the   ${\cal N}=2 $  version of  the theory, with the
adjoint VEV of the form (\ref{vevsonbis}) with $v=m$,    is not sufficient to guarantee  the  unbroken group $ U(N)$   to remain so in the infrared. 
With the standard  ${\cal N}=2 $  embedding 
 the number of  the flavors  in the vector representation  is limited to  $ n_f < 2N-2, $  from the requirement that the original gauge group is
asymptotically free.
       But then   the ``unbroken" 
$SU(N)$ theory    necessarily grows strong in the infrared and breaks itself to an  abelian group   $U(1)^{N-1}$. 
Thus in order to discuss quantum nonabelian monopoles  in a second rank tensor representation of $SU(N)$       one must embed   the
system either   in a  ${\cal N}=4 $    context, or   to consider   different matter content.

No such problem arises  for the system with  $SO(2N) \to  SU(r)  \times  U(1)^{N-r+1}$, with   $r < N$,    where  semiclassical
monopoles in the fundamental as well as in the second rank tensor representations appear.  As long as $r < {n_f \o 2}$,   $n_f < 2 N-2$,  one
easily accomodates the asymptotic freedom of the original theory with   infrared freedom of the subgroup $SU(r)$, by giving the equal bare
masses to all the  hypermultiplets and by choosing the  adjoint scalar VEVs of the form   (\ref{vevsonbis}) with $v=m$.     In the region 
$m \sim \Lambda$,  the semiclassical reasoning only  does not tell which nonabelian monopoles survive in the infrared;   the quantum
analysis of  
\cite{CKKM}    shows that  it is  the monopoles in the fundamental representation of the dual $SU(r)$  that becomes light due to the  quantum
effects.

The situation is a little similar, in the case of  ${\cal N}=2 $   theories with  symmetry  breaking, $SU(N) \rightarrow SU(r) \times SU(s) \times
U(1)^{N-r-s+1} 
$.   For $r$ and $s$ not too large,  the condition for the persistence of quantum monopoles  
    $f_1 \ge    2\, r,$    $f_2  \ge    2 \, s$ can be satisfied in an appropriate vacuum,    where    $f_1$  ($f_2$)  is   the number  of  flavors with
bare    mass
$m_1$  ($m_2$),    
$f_1 + f_2 < n_f$.   All other
$n_f- f_1 - f_2$  flavors   must be given unequal bare masses $m_i$.  Again, the case of maximal nonabelian factors ( $r+s=N$ )     discussed in
the subsection 
\ref{sec:Grassmannian},     is special, in that  the only semiclassical  monopoles are in the  $({\underline r}, {\underline s}^*)$   representation in
this case,     and in that    the    requirement of  persistence of   unbroken 
$SU(r) \times SU(s) $ group,  is not compatible with   the   asymptotic freedom of  the fundamental  $SU(N)$   interactions, in the standard 
${\cal N}=2$  setting.

As we have already mentioned nonabelian monopoles also appear  in  the nontrivial superconformal vacua   of 
$ {\cal N} =2$ models.    Examples include the limiting case   ($r=\left[{n_f \o 2}\right] $)      of the so-called $r$ vacua 
\begin{equation}  SU(N+1),\,\,  SO(2N),    \,\, SO(2N+1)  \,\,  {\hbox {\rm or}} \,\, USp(2N)  \to     SU({n_f \o 2})  \times  
U(1)^{N-{n_f \o 2} +1}
\end{equation}
with equal nonzero  bare quark masses, 
as well as   {\it all }       of  the confining vacua of  $SO(N)$, and $USp(2N)$  theories with vanishing bare quark masses \cite{CKM,CKKM}. 
  The physics in these cases, although
perhaps the most interesting  from the point of view of understanding QCD,  
is complicated by the simultaneous presence  of relatively nonlocal massless monopoles and dyons. 
No local effective Lagrangian description is available,  in general. To
the best of our knowledge, the study of the physical properties of this kind of  systems   is 
still at an exploratory stage  (see for example \cite{AGK}, however).

 \section{Monopole Spectra from Homotopy Groups  \label{sec:homotopy}} 

In this section we attempt to classify topologically inequivalent monopole configurations, leaving the explicit construction 
of the monopoles to  Section \ref{sec:Explicit} and \ref{sec:General}.   A monopole is topologically stable whenever a 2-sphere
 surrounding the monopole supports a nontrivial gauge bundle.  The topology of this bundle is determined entirely by the homotopy type $\pi_1(H)$ of the
transition function $S^1\longrightarrow H$ on the equatorial circle.  This is not the only data that specifies a monopole configuration, in addition there is an
adjoint Higgs field, which is valued in the original gauge group $G$.  Equivalence classes of the Higgs field under the unbroken $H$ gauge symmetry give a
$G/H$-valued function.  The asymptotic value of the Higgs field on the 2-sphere at infinity gives a map $S^2\longrightarrow G/H$ that represents a class
$\pi_2(G/H)$.  Thus any configuration seems to be characterized by pair of topological invariants: $\pi_1(H)$ and $\pi_2(G/H)$.

Here we are interested in only the subset of monopoles which are
finite-energy, regular field configurations of 't Hooft-Polyakov type.
In particular cases, e.g., when the model is embedded in a ${\cal
  N}=2$ theory these monopoles are BPS: they satisfy the linear,
nonabelian Bogomolny equations.  Otherwise, they are solutions only to
quadratic Yang-Mills-scalar coupled field equations.  These equations
allow us to determine up to a constant the profile of the Higgs field
from that of the gauge field and vice versa.  This means that if we
are given only, for example, a class in $\pi_1(H)$ that describes the
gauge field configuration then there is at most one class in
$\pi_2(G/H)$ that describes the Higgs VEV.  Thus each distinct BPS
monopole\footnote{In the following we shall use the term, ``BPS 
  monopoles", having in mind the particularly elegant cases such as
  ${\cal N}=2$ or ${\cal N}=4$ theories, but the whole discussion is
  valid for more general nonsingular solitonlike monopoles. } is
entirely characterized by a class in $\pi_1(H)$.  However there may be
classes in $\pi_1(H)$ that do not correspond to any class in
$\pi_2(G/H)$ and therefore do not correspond to any BPS monopole.
This means that in general BPS monopoles are classified by only a
subset of $\pi_1(H)$, the subset that correspond to elements of
$\pi_2(G/H)$.

So to classify BPS monopoles we need to know what the correspondence is between gauge field configurations $\pi_1(H)$ 
and Higgs field configurations $\pi_2(G/H)$.  This correspondence is determined by the nonabelian Bogomolny equations, but we claim that this is the same
correspondence that arises from the long exact sequence for homotopy groups of fibrations
\begin{equation}
  \label{eq:1}
  \cdots \rightarrow \pi_n (H) \rightarrow \pi_n (G) \rightarrow \pi_n (G/H)
  \rightarrow \pi_{n-1} (H) \rightarrow \cdots .
\end{equation}
Using Poincar\'e's theorem ($\pi_2(G)=0$ for any compact Lie group) and restricting our attention to cases in which $G/H$ is simply connected we find the short exact sequence
\begin{equation}
  \label{eq:1b}
  0 \rightarrow \pi_2 (G/H) \stackrel{f}{\rightarrow} \pi_1 (H) \rightarrow \pi_1 (G) \rightarrow 0.
\end{equation}
The desired correspondence is then the above inclusion $f$ of
$\pi_2(G/H)$ into $\pi_1(H)$, which implies that every class in
$\pi_2(G/H)$ corresponds to a BPS monopole.  On the other hand if $G$
is not simply connected then only one element of $\pi_1(H)$ of every
$|\pi_1(G)|$\ \ (the cardinality of $\pi_1(G)$)\ elements corresponds
to a nonsingular monopole.\footnote{Alternately, ``the condition for a
  nonsingular monopole is that the topological charge is in the kernel
  of the mapping $\pi_1(H) \to \pi_1(G)$" \cite{SC}.  While our
  perspective makes use of the Bogomolny equations to identify the BPS
  monopoles, this perspective uses the topology of the gauge bundle to
  identify the nonsingular monopoles.  In view of Eq.(\ref{equiv})
  these two classification schemes yield identical monopole spectra.}
The others correspond to configurations containing singular Dirac-like
monopoles in the original group $G$, which are enumerated by
$\pi_1(G)$ and do not involve the Higgs field.

Our problem is then reduced to the problem of finding $\pi_2(G/H)$.  The above short homotopy sequence yields the relation
\begin{equation}
  \label{eq:4}
  \pi_1 (G) = \pi_1(H) / \pi_2 (G/H). \label{equiv}
\end{equation} 
This formula, together with an embedding of $H$ into $G$ will always allow us to determine $\pi_2(G/H)$ and thus
 to enumerate the nonabelian monopoles.  As every monopole configuration corresponds to a class in $\pi_1(H)$ that is represented by the transition function
of the gauge bundle, this correspondence will automatically allow us to find the transition functions in the cases below and thus to determine the topologies
of the gauge bundles.  For this purpose we will make extensive use of Table~\ref{centers} which lists the centers and fundamental groups of the relevant
semi-simple Lie groups.

\begin{table}[h]  \label{lista}
\begin{center}  
\begin{tabular}{| c c c    |   }    
   \hline  {\bf G}   & {\bf Center  }       & $\pi_1(G)   $   
\\    \hline    $SU(N+1)$ & $ {\mathbb Z}_{N+1}$  &
${\bf 0}  $ 
   \\  \hline    $ USp(2N)  $ &   ${\mathbb Z}_2  $      & ${\bf 0}  $ 
   \\  \hline $SO(2N+1) $ &  ${\bf 1} $   
        &   $ {\mathbb Z}_{2}$  
\\  \hline $Spin(2N+1) $ &   $ {\mathbb Z}_{2}$ 
        &   ${\bf 0}  $ 
 \\  \hline  $SO(2N)$   &   ${\mathbb Z}_2  $    
       &    $ {\mathbb Z}_{2}$  
\\  \hline  $Spin(4N)$    &  $ {\mathbb Z}_2 \times  {\mathbb Z}_2 $
       &    ${\bf 0}  $ 
\\  \hline  $Spin (4N+2)$   &  ${\mathbb Z}_4  $   
       &    ${\bf 0}  $
   \\
 \hline
\end{tabular}
\caption{{\small  Centers and the fundamental groups of various compact Lie groups  }     }
\label{centers}         
\end{center}   
 \end{table}

\subsection{$SU(N+1 ) \rightarrow SU(N) \times U(1)$}
Let us start by revisiting the case of an $SU(N+1) $ theory
spontaneously broken to $H=U(N)=SU(N)\times U(1)/{\mathbb Z}_{N}$.
The coset space is nothing but the projective sphere ${\mathbb C}P^{N}
= U(N+1)/(U(N)\times U(1))$.  By abuse of notation we will sometimes
omit the quotient, as we have done in the title of this subsection.
This discussion can be easily generalized to, for example, $SU(N+1)
\rightarrow SU(r) \times U(1)^{N+1-r}$.
  
The generator of the unbroken  $U(1)$ gauge group is
\begin{equation}
  \label{eq:3}
  Q = {\rm diag} (1, \cdots, 1,-N).
\end{equation}
Meanwhile the $SU(N)$ consists of $N+1$ by $N+1$ matrices whose top
left $N$ by $N$ submatrix is in $SU(N)$ and whose remaining row and
column consist of all zeros and a single one in the lower-right
corner.  The $N$ elements
\begin{equation}
  e^{2\pi iQ/N} = {\rm diag} (e^{2\pi i/N}, \cdots, e^{2\pi
    i/N},1).\end{equation}
appearing in $U(1)$ are also the ${\mathbb Z}_{N}$ center of the
unbroken $SU(N)$ group.  This is the reason for the ${\mathbb Z}_{N}$
quotient in the definition of $H$.

\begin{figure}[ht]     
\begin{center}
\leavevmode
\epsfxsize 6 cm           
\epsffile{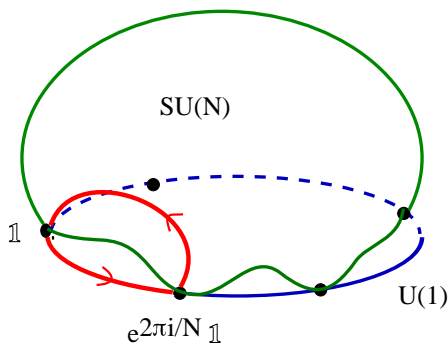}               
\end{center} 
\caption{The smallest closed loop in  $U(N)$ }   
\label{ZN}     
\end{figure}

$G=SU(N+1)$ is simply connected, and so $\pi_2 (SU(N+1)/U(N))$ is equal to $\pi_1(U(N))$, implying that either one classifies nonsingular (BPS in a supersymmetric embedding) monopoles.
To calculate $\pi_1(U(N))$ we use the  exact sequence
\begin{eqnarray}
  \label{eq:5}
  & &0 = \pi_1 ({\mathbb Z}_{N}) \rightarrow \pi_1 (SU(N) \times U(1))
  \rightarrow \pi_1 (U(N)) 
  \nonumber \\
  & & \rightarrow \pi_0
  ({\mathbb Z}_{N}) \rightarrow \pi_0 ((SU(N)\times U(1))/{\mathbb Z}_{N}) = 0.
\end{eqnarray}
By the Kunneth formula $\pi_1 (SU(N) \times U(1)) = \pi_1 (U(1)) = {\mathbb Z}$, while   
$\pi_0 ({\mathbb Z}_{N}) = {\mathbb Z}_{N}$. 
This short exact sequence then yields the equality
\begin{equation}
  \label{eq:6}
  {\mathbb Z}_{N}=\pi_1 (U(N)) / {\mathbb Z},    
\end{equation}
 implying  that $\pi_1 (U(N))$, the group that classifies monopoles, is isomorphic to ${\mathbb Z}$
 possibly times a finite cyclic group whose generator is a loop from the identity to the generator of ${\mathbb Z}_{N}$.  In fact there is no additional
finite cyclic group:    the new generator is just $1/N $ of the old generator of the $\pi_1(U(1))$ (See Fig.\ref{ZN}.)

This point is easy to understand.  In $SU(N)\times U(1)$, the only
non-contractible loops are those that go around the $U(1)$ an integral number of times.
However within $H=U(N)$, there is a loop that goes inside $SU(N)$ from the identity element to the ${\mathbb Z}_{N}$ center,
which can also be regarded as an element of $U(1)$, and then comes back to
the identity inside of the $U(1)$.  If we travel around this loop $N$ times then we have gone around the $U(1)$ one full time, and we have also made a loop in $SU(N)$.  $SU(N)$ is
simply-connected, and so we can deform away the loop that we have made in $SU(N)$, and conclude that $N$ times the generator of $\pi_1(U(N))$ is the generator of $\pi_1(SU(N)\times
U(1))=\pi_1(U(1))$.

We have 
then learned that
$\pi_1 (U(N))$ is isomorphic to ${\mathbb Z}$, yielding precisely one family  of 
BPS monopoles. 

The fact (Eq.(\ref{suncharge})) that the $U(1)$ magnetic charge of the nonabelian monopole is $1/N$ in units of the Dirac quantum for the product theory is due to this $N$ to $1$ embedding
of $\pi_1(SU(N)\times U(1))$ into $\pi_1(U(N))$.  Our monopole has a smaller magnetic charge than any monopole that could exist in a theory with only this $U(1)$ because we have a smaller
loop at
 our disposal with which to construct the transition function of the gauge bundle.  Equivalently the minimal electric charge with respect to the $U(1)$ in the $SU(N+1)\rightarrow
U(N)$ theory is $N$ times the Dirac quantum for the $U(1)$ due to the constraint that all matter must descend from representations of the original $SU(N+1)$.

To make this discussion more concrete, let us present the gauge
transformation (transition function) along the equator of an $S^2$ that
surrounds a monopole \cite{WY}.  The coordinate along the equator is the azimuth
$\phi$, and we have the $SU(N)$ element
\begin{equation}
  \label{eq:7}
  {\rm diag} (e^{-i\phi(N-1)/N},e^{i\phi/N}, \cdots, e^{i\phi/N},
   1)
\end{equation}
and the $U(1)$ element
\begin{equation}
  \label{eq:8}
  {\rm diag}(e^{-i\phi/N}, \cdots, e^{-i\phi/N},
  e^{-i\phi/N}, e^{i\phi}).
\end{equation}
The action of these elements on matter in the fundamental representation
is just the product of the two,
\begin{equation}
  \label{eq:9}
  {\rm diag}(e^{-i\phi},1, \cdots, 1,  e^{i\phi}).
\end{equation}
It clearly is single-valued when one comes back around the equator.
In fact it represents the generator of $\pi_1(U(N))$, as it is just
the loop described above in which one travels to a root of unity in
the $U(1)$ and then returns to the identity inside the $SU(N)$. Thus
we have found the transition function for a monopole configuration
with a single unit of charge.  This monopole is nothing but the
embedding of the usual $SU(2)$ monopole between the first unbroken
color $i=1$ and the abelian part $i=N+1$, {\textit{i.e.}} the
$(1,0,\cdots)$ - monopole. 

One might be tempted to believe that the choice of unbroken color,
that is the choice of loop in $SU(N)$, leads to $N$ independent
monopole solutions.  However the transition functions corresponding to
all possible choices are homotopic to each other, and so the topology
of the bundle is independent of this choice.  The dual gauge
transformations under which our monopoles transform, being continuous,
necessarily preserve the topology of the gauge bundle.  Thus we cannot
find the dimensions of the representations inhabited by monopoles by
merely counting distinct bundles nor by counting homotopically
inequivalent transition functions because all transition functions in
a given multiplet are homotopic.  

The main point is that nonabelian monopoles transform under the dual
gauge group, which in this case is another copy of the original
unbroken gauge group $H=U(N)$.  To determine what representation of
$U(N)$ acts on our monopoles, we decompose the monopoles in terms of
the $U(1)$ and $SU(N)$ actions.  We have seen that the $U(1)$ charge
is the minimum possible value, $1/N$, and so we need only find the
representation of $SU(N)$ under which the monopoles transform.

We claim that in general monopoles transform under some subset of the $k$th symmetric tensor representation of the dual gauge group 
if their transition function represents the element $k\in\Z$ in 
$H$.  In general the subset and therefore the representation is not
determined by the topology,\footnote{While the representation is not
  determined from the gauge bundle's topology alone, it is determined
  by the Lie algebra-valued curvature 2-form of the bundle, which is
  the monopoles magnetic flux. It transforms under the same
  representation as the monopole.} but in this case we have seen that
$k=1$ and so our monopoles transform in a rank one tensor
representation of $SU(N)$.  There are only two such representations,
the fundamental and antifundamental representations.  Thus we may
conclude that the lightest mass monopoles transform in the
(anti-)fundamental representation of $SU(N)$ and have charge $1/N$
under $U(1)$, in short, they transform under the $N$ of $U(N)$.

This construction is consistent with the general construction of
\cite{GNO,EW}, summarized in \ref{sec:General}, according to which
each nonabelian monopole is associated with a root vector
corresponding to a broken generator, and their properties are
determined by the dual of such root, $\alpha^* \equiv \alpha /
\alpha\cdot \alpha$.  The monopoles arising from the breaking $SU(N+1
) \rightarrow SU(N) \times U(1)$ transform in the (anti-)fundamental
representation of the dual $SU(N)$.

\subsection{$SU(N+M) \rightarrow SU(N) \times SU(M) \times U(1)$
\label{sec:Grassmannian}} 

This is a simple generalization of the previous case, yet is it
interesting that the unbroken group involves two non-abelian groups.
The coset group is actually $H = (SU(N) \times SU(M) \times
U(1))/{\mathbb Z}_{k}$ where $k$ is the least common multiple $k =
\mbox{LCM}(N,M)$.  The coset space is nothing but the Grassmannian
manifold $G_{N,N+M}({\mathbb C}) = U(N+M)/(U(N)\times U(M))$.  The
homotopy groups are obtained by
\begin{equation}
  0 = \pi_2 (H) \rightarrow \pi_2(G/H) \rightarrow \pi_1(H) = {\mathbb
  Z} \rightarrow \pi_1(G) = 0.
\end{equation}
Therefore, the monopoles are classified by $\pi_2(G/H) = {\mathbb
Z}$.

The generator of $U(1)$ is
\begin{equation}
  Q = \left( 
    \begin{array}{c|c}
      \frac{1}{N} {\bf 1}_N & \\ \hline
      & -\frac{1}{M} {\bf 1}_M
    \end{array} \right).
\end{equation}
It is clear that $kQ$ has integer eigenvalues by definition, and hence
$e^{2\pi i Q}$ is the generator of ${\mathbb Z}_k$.  The transition
function along the equator is given by $e^{i Q \phi}$ for the $U(1)$,
which is canceled by the center elements of $SU(N)$ and $SU(M)$.

Following the same analysis as in the previous section, the smallest
monopole transforms as $(N, \overline{M})_{1/k}$ under $SU(N) \times SU(M)
\times U(1)$.

\subsection{$SO(N+2) \rightarrow SO(N) \times U(1) $ \label{sec:SOodd}   }

As always, the BPS nonabelian monopoles are classified by
$\pi_2(G/H)=\pi_2(SO(N+2) / SO(N) \times U(1))$.  This homotopy group
may be evaluated using the short exact sequence
\begin{eqnarray}
  \label{eqson}
  & &0 = \pi_2 (SO(N+2))  \rightarrow \pi_2 ({SO(N+2)  /    SO(N) \times U(1) }  )
  \rightarrow
  \nonumber \\
  & & \stackrel{f}{\rightarrow} \pi_1(SO(N) \times U(1))=\Z\times\Z_2 \stackrel{g}{\rightarrow} \pi_1 (SO(N+2))=\Z_2    \rightarrow  0.
\end{eqnarray}
It follows that $ \pi_2 ({SO(N+2) / SO(N) \times U(1) } ) ={\mathbb
  Z}$, and so again we find precisely one family of BPS monopoles.

To find the transition function we use the above map $f$, whose image
is the set of homotopy classes of transition functions of the gauge
bundle.  The surjectivity of $g$ implies that the image of $f$
consists of only half of the elements of $\pi_1(H)$
\begin{equation}
  f(\pi_2 ({SO(N+2)  /    SO(N) \times U(1) } ))=\Z\subset
  \Z\times\Z_2=\pi_1(SO(N) \times U(1)). 
\end{equation}
The transition function representing the nontrivial element of $\Z_2$
is not in the image of $f$ and so describes a singular, Dirac-like
monopole.  In fact it is the usual $\Z_2$ monopole that exists in $SO$
gauge theories even when the symmetry is unbroken \cite{WY}.  The
transition function of the smallest BPS monopole represents the
generator of $\Z\subset\pi_1(H)$ which is the image under $f$ of the
generator of $\pi_2(G/H)$.   

This monopole has a full unit of charge with respect to that of a pure
$U(1)$ monopole because, unlike the previous case, the $U(1)$ is not
quotiented and so the transition function $S^1\rightarrow H$ needs to
wrap the entire $U(1)$ in order to be single-valued.  Indeed we may
explicitly construct the transition function around the equator as
follows.  We can make the $2\pi$ rotation in $SO(N)$ along any
direction, and with no loss of generality we take consider the
rotation in the $(j,k)$ plane.  The unbroken $U(1)$ group is also a
rotation but in the $(N+1, N+2)$ plane.  In other words, all the
action is within an $SO(4) = SU(2) \times SU(2)$ subgroup of
$SO(N+2)$, and the smallest monopole is nothing but the monopole in
one of the $SU(2)$ factors of $SO(4)$.

There are ${}_{N}C_2 = (N-1) N/2 $ choices of $(j,k)$ plane in
$SO(N)$.  Apparently, there are $ (N-1) N/2 $ independent monopole
solutions.  However this argument is too fast as every choice of
$(j,k)$ leads to a homotopically equivalent transition function and so
a topologically equivalent bundle.  Instead we apply the above
proposal that monopoles transform in a $k$th tensor representation of
the dual gauge group if their transition function represents the
element $k$ in the free part of $\pi_1(H)$.  We have seen that for the
lightest monopoles, constructed above, $k=1$ and so these monopoles
transform in the fundamental representation of the dual $SO(N)$ (if
$N$ is even) or $Sp(N-1)$ (if $N$ is odd) gauge group.  We will find
the same result in the next section when we explicitly construct the
monopole solutions.

\subsection{$USp(2N + 2) \rightarrow USp(2N)\times U(1)$\label{sec:USp}}

This example is simpler than the previous as $USp$ groups are simply
connected.  The usual short exact sequence then provides an
isomorphism
\begin{equation}
  \pi_2\left(\frac{USp(2N+2)}{USp(2N)\times
  U(1)}\right)\cong\pi_1(USp(2N)\times U(1))\cong\Z . 
\end{equation}
Again we find a single family of BPS monopoles.  However this time the
simply-connectedness of $USp$ ensures that there are no singular
monopoles and so every transition function in $\pi_1(H)$ yields a
desired gauge bundle.  The mass is determined by the $\pi_1(U(1))$
contribution to the transition function, and so the lightest BPS
monopoles will be those corresponding to the degree one map from the
equator to $U(1)$.

Again the fact that this map is the generator of the fundamental group
implies that these monopoles transform in the fundamental
representation of the dual group, which is $SO(2N+1)$.  The appearance
of a $(2N+1)$-dimensional representation of monopoles constructed
using $USp(2N)$ may appear surprising.  We will see when we explicitly
construct these monopoles in the next section that in addition to the
$2N$ monopoles constructed from the $N$ $SU(2)$'s in $USp(2N)$ whose
Cartan generators generate the Cartan subalgebra of $USp(2N)$, there
is a $(2N+1)$th monopole that is the standard 't Hooft Polyakov
monopole corresponding to the extra unbroken $U(1)$.  In the full
quantum theory these apparently very different types of monopoles are
in fact degenerate and gauge equivalent under the dual $SO(2N+1)$.
This nicely illustrates the importance of distinguishing between the
unbroken gauge symmetry and its dual.

\subsection{$SO(2N) \rightarrow U(N)$\label{sec:SOevenU}}

This breaking is possible, for example, in an ${\cal N}=2$ $SO(2N)$ gauge theory when
the quarks in the vector representation have a degenerate mass which is
precisely canceled by the adjoint VEV.  As this mass approaches zero the full $SO(2N)$ symmetry is restored.

The homotopy exact sequence is now
\begin{eqnarray}
  0 &=& \pi_2 (SO(2N)) \rightarrow \pi_2 (SO(2N)/U(N)) \stackrel{f}{\rightarrow} \pi_1
  (U(N))=\Z \\&&\stackrel{g}{\rightarrow} \pi_1 (SO(2N)) = {\mathbb Z}_2 \rightarrow 0=\pi_1(SO(2N)/U(N)).
\label{twice}\end{eqnarray}
To see that $\pi_1(SO(2N)/U(N))$ vanishes, we will use the fact that our monopoles are all embeddings 
of the standard 't Hooft Polyakov monopoles in broken $SU(2)$ gauge theories.  Thus it suffices to consider the topology of a $SO(4)\rightarrow U(2)$ in which
the monopole is embedded, and so we need only show that $\pi_1(SO(4)/U(2))=0$.  $SO(4)$ is the twisted product of two three-spheres, and the action of
$SU(2)$ on each 3-sphere is given by the identification of the 3-sphere with $SU(2)$ and the group multiplication in $SU(2)$.  We may then quotient $SO(4)$ by
$U(2)$ in two steps, first we quotient by the action of $SU(2)$ and then by the action of $U(1)$. 
 As the $SU(2)$ acts diagonally on the two 3-spheres, $SO(4)/SU(2)$ is $SU(2)/Z_2=SO(3)$, which is topologically a circle bundle over a 2-sphere with Chern
 class equal to two.  The quotient of the total space of this bundle by its fiber, a circle, is just the base space of the bundle, which is a 2-sphere. 
 The 2-sphere
is simply-connected and so $\pi_1(SO(4)/U(2))=\pi_1(S^2)=0$ as claimed.  

The surjectivity of $g:\Z\rightarrow\Z_2$ implies that the image of $f$ is the set of even integers.  
$f$ must be one to one and so its domain, $\pi_2(SO(2N)/U(N))$ must be the set of integers.  This 
is consistent with the above topological description of
$SO(2N)/U(N)$ as the 2-sphere, as $\pi_2(SO(2N)/U(N))=\pi_2(S^2)=\Z$.  Every element of $\pi_1(U(N))=\Z$ describes 
a gauge bundle, but as only the even
elements are in the image of $f$, only the even elements describe BPS monopoles.  The odd elements are transition
 functions of singular configurations that
include the singular $\Z_2$ monopoles of the original $SO$ gauge theory.  Due to the $N$ to $1$ embedding of $U(1)$ 
in $U(N)$, the lowest possible monopole
charge is $1/N$ of the Dirac quantum corresponding to the element $1\in\pi_1(U(N))$.  Therefore the minimum 
possible charge of a BPS monopole is $2/N$,
corresponding to the smallest even element $2\in\pi_1(U(N))$.

Using our proposed relation between elements of the fundamental group and representations we then conclude 
that, in addition to the above $U(1)$ charge these monopoles transform under some second rank tensor representation 
of the dual $SU(N)$ gauge symmetry. 
However topological considerations alone do not allow us to determine which second rank tensor representation yields a BPS monopole.  The answer to this
question will need to await the next section, in which the monopoles are explicitly constructed.

\subsection{$SO(2N+1) \rightarrow U(N )$\label{sec:SOoddU}}

The analysis in this case proceeds identically to the preceding case.  The verification that $\pi_1(SO(2N+1)/U(N))=0$ again reduces to 
a computation of $SO(4)/U(2)$ using the fact that the monopole solution is the embedding of an $SU(2)$ monopole (or more accurately perhaps an $SO(3)$
monopole) into some $SO(4)\rightarrow U(2)$ subsector.  And so we again conclude that there are BPS nonabelian monopoles with charge $2/N$ under the
unbroken $U(1)$ that transform under an unknown second rank tensor representation of the dual $SU(N)$.  The fact that the topology does not tell 
us which second
rank tensor to use is highlighted by the fact that, as we will see, despite the identical homotopy groups in the two examples the corresponding monopoles
transform under different second rank tensor representations of the dual $SU(N)$.

\subsection{$USp(2N) \rightarrow U(N)$\label{sec:UspU}}

This breaking is possible, for example, in an ${\cal N}=2$
$USp(2N)$ theory when the $N$ quarks have a degenerate mass which is
precisely canceled by the adjoint VEV.  The fundamental representation
of $USp(2N)$ decomposes as $2N = (N, +1) \oplus (\overline{N}, -1)$.
The center of $SU(N)$ can be regarded as an element of $U(1)$, and
hence the quotient by ${\mathbb Z}_N$.  

$USp(2N)$ is simply connected and so every transition function in
$\pi_1(SU(N))$ is the image of some Higgs field in
$\pi_2(USp(2N)/SU(N))$.  Thus for every allowed charge in
$\pi_1(SU(N))=\Z$ there is a solution to the Bogomolny equations
yielding a BPS monopole configuration The smallest such charge, in
units of the Dirac quantum of the $U(1)$, is then $1/N$.  The
corresponding transition function is a representative of the generator
of $\pi_1(U(N))$.  This may be written as a product of generators of
the fundamental groups of $SU(N)$ and $U(1)$, such as ${\rm diag}(e^{i
  \phi/N}, \cdots, e^{i \phi/N}, e^{-i (N-1)\phi/N}) \in SU(N)$ and
$e^{-i\phi/N} \in U(1)$ which multiply to $$
{\rm diag} (1, \cdots, 1,
e^{-i \phi}) \in U(N). $$

Looking at the whole fundamental representation of $USp(2N)$, one can
see that it is a monopole embedded in $SU(2) \simeq USp(2) \subset
USp(2N)$.  The monopoles therefore transform as the fundamental
representation of the $SU(N)$ group, as follows, for example, from the
fact that the transition function represents $1\in\pi_1(H)$.

\subsection{$SO(N+2M) \rightarrow SO(N) \times U(M)$}

The breaking is possible, for example, in an ${\cal N}=2$ $SO(N+2M)$
theory when $M$ quarks have a degenerate mass which is precisely
canceled by the adjoint VEV.  The fundamental representation of
$SO(N+2M)$ decomposes as $N+2M = (N,1) \oplus (1,M) \oplus
(1,\overline{M})$.  

Combining analyses in Sec.~\ref{sec:SOodd}, \ref{sec:SOevenU},
\ref{sec:SOoddU}, we find that the monopoles are classified according
to $\pi_2 (SO(N+2M)/(SO(N)\times U(M)) = \pi_1(U(M)) = {\mathbb Z}$.
When $N$ is even, the monopoles transform under the dual $SO(N) \times
U(M)$ group as $(N, M)_{1/M} \oplus (1,
\frac{M(M-1)}{2})_{2/M}$, where the latter may decay into the
pair of the former.  When $N$ is odd, the monopoles transform under
the dual $USp(N-1) \times U(M)$ as $(N-1,M)_{1/M} \oplus
(1,\frac{M(M+1)}{2})_{2/M}$, where the latter may decay into the pair
of the former. 

\subsection{$USp(2N+2M) \rightarrow USp(2N) \times U(M)$}

The breaking is possible, for example, in an ${\cal N}=2$ $USp(2N+2M)$
theory when $M$ quarks have a degenerate mass which is precisely
canceled by the adjoint VEV.  The fundamental representation of
$USp(2N+2M)$ decomposes as $N+2M = (2N,1) \oplus (1,M) \oplus
(1,\overline{M})$.  

Combining analyses in Sec.~\ref{sec:USp} and \ref{sec:UspU}, we find
that the monopoles are classified according to $\pi_2
(USp(2N+2M)/(USp(2N)\times U(M)) = \pi_1(U(M)) = {\mathbb Z}$.  The
monopoles transform under the dual $SO(2N+1) \times U(M)$ group as
$(2N+1, M)_{1/M} \oplus (1, \frac{M(M-1)}{2})_{2/M}$, where the latter
may decay into the pair of the former.

\section{Explicit Constructions of BPS
Non\-abelian Mono\-poles \label{sec:Explicit}  }

In this section  we construct  the lightest    nonabelian monopole solutions  by simply identifying various 
minimally embedded $SU(2)$  subgroups \cite{BCGW} which are broken to $U(1)$, and  then by embedding the 't Hooft-Polyakov  monopoles into this $SU(2)$ and
adding a constant term to $\phi$  so that  in a given spacial direction it takes the standard, prescribed form.  The results agree in all cases with the general
formulae given in \ref{sec:General} and with the classifications of Section~\ref{sec:homotopy}.     The $SU(N+1)$ case has already been discussed in
Section~\ref{sec:ordinary}, and so below we consider the $SO(N)$ and $USp(2N)$ examples.  

In the previous section we characterized monopoles by $\pi_1(H)$, whose representatives are transition functions of the gauge bundles of BPS monopoles.  In this section we will find the
adjoint scalar profiles, whose gauge equivalence classes represent elements of $\pi_2(G/H)$.  The interested reader need only substitute these results into the nonabelian Bogomolny equations to
find the corresponding gauge potentials.

The explicit solutions of the adjoint scalars will allow us to determine in each case the representation of the dual gauge group under which the nonabelian monopoles transform.  The relevant
pairs of dual groups  are listed in Table~\ref{tabdual}.
{\large
\begin{table}[h]
\begin{center}
\begin{tabular}{c  c   c}
\hline
$SU(N)/{\mathbb Z}_N       $        &
  $\Longleftrightarrow$                 &    $SU(N)     $
         \\
  $ SO(2N)  $     &   $\Longleftrightarrow$    &   $SO(2N)
$       \\
  $ SO(2N+1)  $     &   $\Longleftrightarrow$     &
  $USp(2N) $       \\ \hline
\end{tabular}   
\caption{Some examples of dual pairs of groups}
\label{tabdual}      
\end{center}  
\end{table}}

\subsection{$SO(2N+3) \rightarrow SO(2N+1) \times U(1)$} \label{exso}
An adjoint scalar VEV of the form
{\small\begin{equation}    \phi =
 i \left( \begin{array} {ccc|cc}  0     &  0  &  \ldots &
   0 & 0
  \\  0   & 0    & \ldots   & 0  & 0
 \\  \vdots  &\vdots& \ddots & \vdots & \vdots
\\ \hline 0  &0  & \ldots    & 0 & v
\\ 0 & 0 & \ldots & -v & 0 \end{array} \right) =
i \left( \begin{array} {c|cc}    {\bf 0}_{(2N+1)\times
(2N+1)}     &  &
  \\ \hline  & 0 & v
\\ & -v & 0 \end{array} \right).
\label {vevson}\end{equation}
}breaks an $SO(2N+3)$ gauge symmetry to $SO(2N+1)\times U(1)$.
We will use two minimally-embedded $SU(2)$ subgroups:
{\small\begin{equation}
S_1=\frac{i}{2} \left( \begin{array} {ccc|cc}
    \ddots & \vdots & \vdots & \vdots & \vdots
\\  \ldots &   0     &   0     &   1     & 0
\\  \ldots &    0    &   0     &   0    &  -1
\\ \hline \ldots &    -1    &  0    &    0    &  0
\\  \ldots &   0   &    1    &    0    &  0
 \end{array} \right), \quad
S_2=\frac{i}{2} \left( \begin{array} {ccc|cc}
    \ddots & \vdots & \vdots & \vdots & \vdots
\\  \ldots &   0     &   0     &   0     & 1
\\  \ldots &    0    &   0     &   1    &  0
\\ \hline \ldots &    0    &  -1    &    0    &  0
\\  \ldots &   -1   &    0    &    0    &  0
 \end{array} \right),    \label{sua1}
\end{equation}
\begin{equation}
S_3=\frac{i}{2} \left( \begin{array} {ccc|cc}
    \ddots & \vdots & \vdots & \vdots & \vdots
\\  \ldots &   0     &   1     &   0     & 0
\\  \ldots &    -1    &   0     &   0    &  0
\\ \hline \ldots &    0    &  0    &    0    &  1
\\  \ldots &   0   &    0    &    -1    &  0
\end{array} \right). \label{sua2} \end{equation}
\begin{equation}
\tilde{S}_1=\frac{i}{2} \left( \begin{array} {ccc|cc}
    \ddots & \vdots & \vdots & \vdots & \vdots
\\  \ldots &   0     &   0     &   -1     & 0
\\  \ldots &    0    &   0     &   0    &  -1
\\ \hline \ldots &    1    &  0    &    0    &  0
\\  \ldots &   0   &    1    &    0    &  0
\end{array} \right), \quad
\tilde{S}_2=\frac{i}{2} \left( \begin{array} {ccc|cc}
    \ddots & \vdots & \vdots & \vdots & \vdots
\\  \ldots &   0     &   0     &   0     & -1
\\  \ldots &    0    &   0     &   1    &  0
\\ \hline \ldots &    0    &  -1    &    0    &  0
\\  \ldots &   1   &    0    &    0    &  0
\end{array} \right),    \label{sub1}    \end{equation}
\begin{equation}
\tilde{S}_3=\frac{i}{2} \left( \begin{array} {ccc|cc}
    \ddots & \vdots & \vdots & \vdots & \vdots
\\  \ldots &   0     &   -1     &   0     & 0
\\  \ldots &    1    &   0     &   0    &  0
\\ \hline \ldots &    0    &  0    &    0    &  1
\\  \ldots &   0   &    0    &    -1    &  0
\end{array} \right). \label{sub2} \end{equation} }

We may create a nonabelian monopole using the first $SU(2)$, in which case the adjoint scalar profile is:
{\small   \begin{equation}
\phi=
\frac{i}{2} \left( \begin{array} {ccc|cc}
    \ddots & \vdots & \vdots & \vdots & \vdots
\\  \ldots &   0     &   -v     &   0     & 0
\\  \ldots &   v    &   0     &   0    &  0
\\ \hline \ldots &    0    &  0    &    0    &  v
\\  \ldots &   0   &    0    &    -v    &  0
\end{array} \right)
- v  (\vec{S} \cdot \widehat{r}) \phi(r)
;\end{equation}  }
or by using the second $SU(2)$:
 {\small   \begin{equation}
\phi=
 \frac{i}{2} \left( \begin{array} {ccc|cc}
    \ddots & \vdots & \vdots & \vdots & \vdots
\\  \ldots &   0     &   v     &   0     & 0
\\  \ldots &   -v    &   0     &   0    &  0
\\ \hline \ldots &    0    &  0    &    0    &  v
\\  \ldots &   0   &    0    &    -v    &  0
\end{array} \right)
- v  (\vec{\tilde{S}} \cdot \widehat{r}) \phi(r).
\end{equation}  }
The two   types of monopole have  the same  mass:
\begin{equation} M= \frac{4 \pi v}{g }. \end{equation}
As  there are  $(2N+1)$ rows of broken group generators,
one  can construct in this way $2N$
independent  minimally-embedded  monopoles.
For the last row one is   forced to use the non-minimal
$SO(3)$ embedding:
{\small   \begin{equation}
S_1= i \left( \begin{array} {cc|cc}
    \ddots & \vdots & \vdots & \vdots
\\  \ldots &   0     &   -1     &   0
\\ \hline \ldots &    1    &   0     &   0
\\  \ldots &    0    &  0    &    0
\end{array} \right), \qquad
S_2= i \left( \begin{array} {cc|cc}
    \ddots & \vdots & \vdots & \vdots
\\  \ldots &   0     &   0     &   -1
\\ \hline \ldots &    0    &   0     &   0
\\  \ldots &    1    &  0    &    0
\end{array} \right),
\end{equation}
\begin{equation}
S_3= i \left( \begin{array} {cc|cc}
    \ddots & \vdots & \vdots & \vdots
\\  \ldots &   0     &   0     &   0
\\ \hline \ldots &    0    &   0     &   1
\\  \ldots &    0    &  -1    &    0
\end{array} \right);     \qquad
\phi=
 v  (\vec{S} \cdot \widehat{r}) \phi(r),
\end{equation}    
}but   this is     a non-minimal monopole with mass
\begin{equation} M= \frac{8 \pi v}{g}, \end{equation}
twice that of the minimal ones.
To summarize,   there are
  $2N$ degenerate minimally embedded monopoles.
We note again that  no irreducible representation of the
gauge group  $ SO(2N+1) $
has such a multiplicity, but this is precisely the multiplicity of the fundamental  representation of the {\it
dual}
group,   $USp(2N)$.

\subsection{$SO(2N+2) \rightarrow SO(2N) \times U(1)$}

This breaking pattern may be achieved with the Higgs VEV:
{\small   \begin{equation}    \phi =
i \left( \begin{array} {ccc|cc}  0     &  0  &  \ldots &
   0
& 0
  \\  0   & 0    & \ldots   & 0  & 0
 \\  \vdots  &\vdots& \ddots & \vdots & \vdots
\\ \hline 0  &0  & \ldots    & 0 & v
\\ 0 & 0 & \ldots & -v & 0 \end{array} \right) =
 i \left( \begin{array} {c|cc}  0 \cdot {\bf 1}_{(2N)\times
(2N)}     &  &
  \\ \hline   & 0 & v
\\ & -v & 0 \end{array} \right).
\end{equation}}We use minimally-embedded $SU(2)$ subgroups as
in the previous case.
The formulas for $S$,$\tilde{S}$,$\phi$ are the same.
We find $2 N$ monopoles with mass:
\begin{equation} M= \frac{4 \pi v}{g}. \end{equation}
This is the right number of degenerate degrees of freedom
for a monopole in the vector representation of the
dual $SO(2N)$ magnetic group.

\subsection{$USp(2N+2) \rightarrow USp(2N) \times U(1)$}

The generic $USp(2N)$ matrix in the fundamental
representation
has the following form:
\begin{equation} Z= \pmatrix{ {\mathbb A} & {\mathbb B } \cr
{\mathbb B }^* & -{\mathbb A}^t}, \end{equation}
with ${\mathbb A}^{\dagger} = {\mathbb A}$ and ${\mathbb
B}^t={\mathbb B }$.
The symmetry breaking field is:
{\footnotesize
\begin{equation}    \phi =
 \left( \begin{array} {ccc|ccc}  0    & \ldots   & 0  & 0 &
\ldots & 0
 \\  \vdots& \ddots & \vdots & \vdots& \ddots & \vdots
\\   0   &  \ldots  & v  & 0  & \ldots & 0 \\
\hline
    0 & \ldots & 0 & 0       & \ldots   & 0
 \\  \vdots& \ddots & \vdots   &\vdots& \ddots & \vdots
\\  0       & \ldots   & 0      & 0   &  \ldots  & -v \\
\end{array} \right)
 = \left( \begin{array} {cc|cc} 0 \cdot {\bf 1}_{N\times N}
    &   &  &
  \cr     &  v & &
\cr  \hline &  & 0 \cdot {\bf 1}_{N\times N} &
\cr  &  &  & -v   \end{array} \right).  \end{equation} }
There are  three kinds of monopoles:  

\subsubsection{First type}
We use the $SU(2)$ subgroup with ${\mathbb B}=0$.
{\footnotesize
\begin{equation} S_1=\frac{1}{2} \left( \begin{array} {ccc|ccc}
  \ddots    &    &   &   &  &
  \\     & 0    & 1   &   & 0 & 0
 \\    & 1 & 0 &  & 0 & 0
\\  \hline     &    &    & \ddots  &  &
 \\  & 0  &0  & & 0     &  -1
  \\   & 0 & 0 &  & -1   & 0
  \end{array} \right) ;\quad   
S_2=\frac{1}{2} \left( \begin{array} {ccc|ccc}
 \ddots    &    &   &   &  &
  \\     & 0    & -i   &   & 0 & 0
 \\    & i & 0 &  & 0 & 0
\\ \hline      &    &    & \ddots  &  &
 \\  & 0 & 0 &  & 0     &  i
  \\   & 0 & 0 &  &  -i   & 0
 \end{array} \right), 
 \end{equation}
\begin{equation} S_3=\frac{1}{2} \left( \begin{array} {ccc|ccc}
  \ddots    &    &   &   &  &
  \\     & 1    & 0   &   & 0 & 0
 \\    & 0 & -1 &  & 0 & 0
\\ \hline      &    &    & \ddots  &  &
 \\  & 0 & 0 &  &   -1   &  0
  \\   & 0 & 0 &  & 0   & 1
  \end{array} \right).    \end{equation} }
to construct the adjoint scalar profile
{\small
\begin{equation}
\phi=\left( \begin{array} {ccc|ccc}
  \ddots    &    &   &   &  &
  \\     & v/2    & 0   &   & 0 & 0
 \\    & 0 & v/2 &  & 0 & 0
\\  \hline    &    &    & \ddots  &  &
 \\  & 0 & 0 &  &   -v/2   &  0
  \\   & 0 & 0 &  & 0   & -v/2
   \end{array} \right)
-v  \, (\vec{S} \cdot \widehat{r}) \, \phi(r).
\end{equation}  } 
There are $N$ monopoles of this kind, each with a mass of $M= \frac{4 \pi v}{g}$.

\subsubsection{Second type }
Alternately we may use the embedding
{\small  
\begin{equation} S_1=\frac{1}{2} \left( \begin{array} {ccc|ccc}
  \ddots    &    &   &   &  &
  \\     & 0    & 0   &   & 0 & 1
 \\    & 0 & 0 &  & 1 & 0
\\  \hline    &    &    & \ddots  &  &
 \\  & 0 & 1 &  & 0     &  0
  \\   & 1 & 0 &  & 0   & 0
   \end{array} \right);\quad  
S_2=\frac{1}{2} \left( \begin{array} {ccc|ccc}
  \ddots    &    &   &   &  &
  \\     & 0    & 0   &   & 0 & i
 \\    & 0 & 0 &  &   i & 0
\\ \hline      &    &    & \ddots  &  &
 \\  & 0 & -i &  & 0     &  0
  \\   &  - i & 0 &  & 0   & 0
   \end{array} \right);
 \end{equation}
\begin{equation}
S_3=\frac{1}{2} \left( \begin{array} {ccc|ccc}
 \ddots    &    &   &   &  &
  \\     & -1    & 0   &   & 0 & 0
 \\    & 0 & -1 &  & 0 & 0
\\ \hline     &    &    & \ddots  &  &
 \\  & 0 & 0 &  & 1     &  0
  \\   & 0 & 0 &  & 0   & 1
   \end{array} \right)  
 \end{equation}  
}leading to an adjoint scalar profile
{\small 
\begin{equation}
\phi=\left( \begin{array} {ccc|ccc}
  \ddots    &    &   &   &  &
  \\     & - v/2    & 0   &   & 0 & 0
 \\    & 0 & v/2 &  & 0 & 0
\\  \hline    &    &    & \ddots  &  &
 \\  & 0 & 0 &  &   v/2   &  0
  \\   & 0 & 0 &  & 0   & -v/2
   \end{array} \right)
-v (\vec{S} \cdot \widehat{r}) \phi(r). 
\end{equation}
}There are again  $N$ monopoles of this kind, each with $ M= \frac{4 \pi v}{g}$.

\subsubsection{Third type}
The last monopole lies  in the $SU(2)$  subgroup
{\footnotesize 
\begin{equation} S_1=\frac{1}{2} \left( \begin{array} {cc|cc}
  \ddots    &    &   &
  \\     & 0    &  & 1
 \\ \hline   &  & \ddots &
\\      & 1   &    &  0
   \end{array} \right);\quad  
S_2=\frac{1}{2} \left( \begin{array} {cc|cc}
  \ddots    &    &   &
  \\     & 0    &  & - i
 \\ \hline    &  & \ddots &
\\      &   i   &    &  0
   \end{array} \right);
\quad
 S_3=\frac{1}{2} \left( \begin{array} {cc|cc}
  \ddots    &    &   &
  \\     & 1    &  & 0
 \\ \hline    &  & \ddots &
\\      & 0   &    &  -1
  \end{array} \right).
\end{equation}  
}There  is only one monopole of this type, 
\begin{equation}
\phi=
2 \, v  \, (\vec{S} \cdot \widehat{r}) \, \phi(r),
\qquad    M= \frac{4 \pi v}{g }.   \end{equation}
To summarize, we have found  $2N+1$ degenerate monopoles states in the system  $USp(2N+2) \rightarrow USp(2N) \times U(1)$.
This is exactly the  number of degrees of freedom for
a monopole in the vector representation of the
dual $SO(2N+1)$ magnetic group.  This case, like that of Subsec.~\ref{exso}, nicely
illustrates the fact that soliton states of
$USp(2N) $ gauge-invariant system  appear with multiplicity
appropriate for the dual group -
$SO(2N+1)$ in this case -  and not  with multiplicities of
any  irreducible representation  of the original group $USp(2N)$.

\subsection{$SO(2N) \rightarrow U(N)$  \label{sec:sonexp1}}
This symmetry breaking pattern is characteristic of the adjoint scalar profile
{\small   \begin{equation}    \phi =
 i \left( \begin{array} {cc|cc|cc}  0     &  v  &  \ldots &
\ldots &  0 & 0
  \\  -v   & 0    & \ldots & \ldots   & 0  & 0
 \\ \hline \vdots  &\vdots& \ddots & & \vdots & \vdots
 \\  \vdots  & \vdots & & \ddots & \vdots & \vdots
\\ \hline 0  &0  & \ldots  & \ldots  & 0 & v
\\ 0 & 0 & \ldots & \ldots & -v & 0 \end{array} \right) .
\label{vevsonbis}\end{equation}
}Consider first of all the  $SO(4) \rightarrow U(2)$ case.
We use the basis $S_i,\tilde{S}_i$ in Eq.
(\ref{sua1})-(\ref{sub2});
these two $SU(2)$ subgroups are orthogonal and commute with each other.  
As the $\tilde{S}_i$ subgroup is actually
unbroken
we can build a monopole using  the $S_i$ generators only:
\begin{equation}  \phi=2v  (\vec{S} \cdot \widehat{r})
\phi(r),  \end{equation}
with a mass 
\begin{equation} M= \frac{8 \pi v}{g}. \end{equation}
In the general $SO(2N) \rightarrow U(N)$ case, we find $\,\, \frac{N(N-1)}{2}\,\,$ degenerate  monopoles,
which can be interpreted as  an antisymmetric 2-tensor
 representation of the (dual)  $SU(N).$

\subsection{$SO(2N+1) \rightarrow U(N)$ \label{sec:sonexp2}}  
This symmetry breaking pattern results from the adjoint Higgs VEV
{\small
\begin{equation}    \phi =
 i \left( \begin{array} {cc|cc|cc|c}  0     &  v  &  \ldots
& \ldots &  0 & 0   &0
  \\  -v   & 0    & \ldots & \ldots   & 0  & 0   & 0
 \\ \hline \vdots  &\vdots& \ddots & & \vdots & \vdots   &
\vdots
 \\  \vdots  & \vdots & & \ddots & \vdots & \vdots  &
\vdots
\\ \hline 0  &0  & \ldots  & \ldots  & 0 & v   & 0
\\ 0 & 0 & \ldots & \ldots & -v & 0 & 0
\\ \hline 0 & 0& \ldots & \ldots & 0 & 0 & 0
 \end{array} \right) .
\end{equation} 
}In this case we may again construct all $\frac{N(N-1)}{2}$  monopoles
found above in the $  SO(2N)  \rightarrow  
U(N)$ system.
But there are another $N$ monopoles corresponding to the following embeddings
{\small
\begin{equation}     S_1 =
 i \left( \begin{array} {c|cc|c}
 \ddots & \vdots & \vdots   & \vdots
\\ \hline \ldots  & 0 & 0   & 0
\\   \ldots & 0 & 0 & 1
\\  \hline \ldots & 0 & -1 & 0
 \end{array} \right), \qquad  
  S_2 =
 i \left( \begin{array} {c|cc|c}
 \ddots & \vdots & \vdots   & \vdots
\\ \hline \ldots  & 0 & 0   & -1
\\   \ldots & 0 & 0 & 0
\\  \hline \ldots & 1 & 0 & 0
 \end{array} \right), \end{equation}
 \begin{equation} S_3 =
 i \left( \begin{array} {c|cc|c}
 \ddots & \vdots & \vdots   & \vdots
\\ \hline \ldots  & 0 & 1   & 0
\\   \ldots & -1 & 0 & 0
\\  \hline \ldots & 0 & 0 & 0
 \end{array} \right) .
\end{equation}
}The scalar profile is then
{\small 
\begin{equation}    \phi =
 i \left( \begin{array} {cc|cc|cc|c}  0     &  v  &  \ldots
& \ldots &  0 & 0   &0
  \\  -v   & 0    & \ldots & \ldots   & 0  & 0   & 0
 \\ \hline \vdots  &\vdots& \ddots & & \vdots & \vdots   &
\vdots
 \\  \vdots  & \vdots & & \ddots & \vdots & \vdots  &
\vdots
\\ \hline 0  &0  & \ldots  & \ldots  & 0 & 0   & 0
\\ 0 & 0 & \ldots & \ldots & 0 & 0 & 0
\\ \hline 0 & 0& \ldots & \ldots & 0 & 0 & 0
 \end{array} \right) +
  v  (\vec{S} \cdot \widehat{r}) \phi(r),
\end{equation}  
}and the mass :
 \begin{equation} M= \frac{8 \pi v}{g}. \end{equation}
 We then find in all  $\frac{N(N+1)}{2}$ degenerate monopoles
 which transform in the symmetric 2-tensor  representation
of $SU(N).$

 \subsection{$USp(2N) \rightarrow U(N)$\label{sec:USpU}}
This gauge symmetry breaking pattern results from the Higgs VEV
   \begin{equation}    \phi
 = \left( \begin{array} {c|c} v \cdot {\bf 1}_{N\times N}
    &
  \cr    \hline & -v \cdot {\bf 1}_{N\times N}
  \end{array} \right).
\end{equation}
In this case we describe the two least massive species of monopole,
one of which transforms in the fundamental
representation of the dual $SU(n)$
and the other     in the antisymmetric 2-tensor
representation.

\subsubsection{Fundamental  Monopole}
We take the following $SU(2)$ subgroups:
{\footnotesize  \begin{equation} S_1=\frac{1}{2}  \left( \begin{array} {ccc|ccc}  0
   & \ldots   & 0  & 1 &
\ldots & 0
 \\  \vdots& \ddots & \vdots & \vdots& \ddots & \vdots
\\   0   &  \ldots  & 0  & 0  & \ldots & 0 \\
\hline
    1 & \ldots & 0 & 0       & \ldots   & 0
 \\  \vdots& \ddots & \vdots   &\vdots& \ddots & \vdots
\\  0       & \ldots   & 0      & 0   &  \ldots  & 0 \\
\end{array} \right) ;\quad
S_2=\frac{1}{2} \left( \begin{array} {ccc|ccc}  0    &
\ldots   & 0  & -i &
\ldots & 0
 \\  \vdots& \ddots & \vdots & \vdots& \ddots & \vdots
\\   0   &  \ldots  & 0  & 0  & \ldots & 0 \\
\hline
     i & \ldots & 0 & 0       & \ldots   & 0
 \\  \vdots& \ddots & \vdots   &\vdots& \ddots & \vdots
\\  0       & \ldots   & 0      & 0   &  \ldots  & 0 \\
\end{array} \right) ;
 \end{equation}
\begin{equation} S_3=\frac{1}{2} \left( \begin{array} {ccc|ccc}  1
   & \ldots   & 0  & 0 &
\ldots & 0
 \\  \vdots& \ddots & \vdots & \vdots& \ddots & \vdots
\\   0   &  \ldots  & 0  & 0  & \ldots & 0 \\
\hline
    0 & \ldots & 0 & -1       & \ldots   & 0
 \\  \vdots& \ddots & \vdots   &\vdots& \ddots & \vdots
\\  0       & \ldots   & 0      & 0   &  \ldots  & 0 \\
\end{array} \right).
\end{equation}  }
The resulting Higgs VEV is
{\footnotesize
\begin{equation}
\phi=
 \left( \begin{array} {cccc|cccc} 0   & \ldots & \ldots   &
0  & 0 &
\ldots & \ldots & 0
 \\  \vdots & v &  & \vdots & \vdots & \ddots &  & \vdots
 \\  \vdots& & \ddots & \vdots & \vdots& & \ddots & \vdots
\\   0   &  \ldots & \dots  & v  & 0  & \ldots & \ldots & 0
\\
\hline
    0 & \ldots  & \ldots & 0 & 0       & \ldots & \ldots  &
0
 \\  \vdots& \ddots & & \vdots   &\vdots& -v  & & \vdots
 \\  \vdots& & \ddots & \vdots   &\vdots& & \ddots & \vdots
\\  0       & \ldots & \ldots   & 0      & 0   &  \ldots &
\ldots & -v \\
\end{array} \right)
+2v  (\vec{S} \cdot \widehat{r}) \,  \phi(r).
\end{equation}
}These embeddings describe $N$ monopoles with the same mass $M= \frac{8 \pi v}{g}$.  
They transform in the fundamental representation of the dual $SU(N)$.  

\subsubsection{Antisymmetric 2-Tensor Monopole}
If instead we consider the following $SU(2)$ subgroups:

\begin{equation} S_1=   \pmatrix { {\bf 0} &  {\bf A}_1 \cr   {\bf A}_1^*  & {\bf 0}}, \quad
S_2=   \pmatrix { {\bf 0} &  {\bf A}_2 \cr   {\bf A}_2^*  & {\bf 0}}, \quad
    S_3=   \pmatrix { {\bf B}_3 &  {\bf 0} \cr   {\bf 0 }  & - {\bf B}_3^T}, \end{equation}
{\small
\begin{equation}  {\bf A}_1=
\frac{1}{2} \left( \begin{array} {ccccc}
 0 & \ldots & 1 & \ldots &0
 \\
 \vdots& \ddots & & &
\vdots
 \\ 1 & &\ddots & & \vdots
 \\   \vdots & & & \ddots &
\vdots
\\   0  & \ldots &
\ldots &\ldots& 0   \\
\end{array} \right)  ; \qquad
 {\bf A}_2=
\frac{1}{2} \left( \begin{array} {ccccc}
 0 & \ldots & -i & \ldots &0
 \\
 \vdots& \ddots & & &
\vdots
 \\ - i & &\ddots & & \vdots
 \\   \vdots & & & \ddots &
\vdots
\\   0  & \ldots &
\ldots &\ldots& 0   \\
\end{array} \right)  ; \end{equation}
\begin{equation}
 {\bf B}_3=
\frac{1}{2} \left( \begin{array} {ccccc}
 1    & \ldots  & \ldots
& \ldots & 0
 \\  \vdots& \ddots & & & \vdots
 \\  \vdots & & 1 & & \vdots
 \\   \vdots & & & \ddots &\vdots
\\   0   &  \ldots & \ldots &\ldots & 0   \\
\end{array} \right).
\end{equation}    
}then we may construct the solutions
\begin{equation}
\phi=  \pmatrix { {\bf B}_{\phi}   &  {\bf 0} \cr   {\bf 0 }  & - {\bf B}_{\phi}^T}
+2 v  (\vec{S} \cdot \widehat{r}) \phi(r)
\end{equation}
where
{\small    \begin{equation}   {\bf B}_{\phi}  =    \left( \begin{array} {ccccc}  0     & \ldots
 & \ldots
& \ldots & 0
 \\  \vdots& v & & & \vdots
 \\  \vdots & & 0 & & \vdots
 \\   \vdots & & & v &\vdots
\\   0   &  \ldots & \ldots &\ldots & v  \\
\end{array} \right).
\end{equation}
}These solutions describe $\frac{N(N-1)}{2}$ degenerate monopoles that transform in the antisymmetric tensor representation of the dual $SU(N)$. Their mass
is twice the mass of the fundamental monopole:
\begin{equation} M= \frac{16 \pi v}{g }.  \end{equation}

\subsection{Others}

Other breaking patterns discussed in the previous section, $SU(N+M)
\rightarrow SU(N) \times SU(M) \times U(1)/{\mathbb Z}_k$, $SO(N+2M)
\rightarrow SO(N) \times U(M)$, and $USp(2N+2M) \rightarrow USp(2N)
\times U(M)$, can be discussed basically by fixing the index of the
latter gauge factor and carry through the same analysis as those in
the similar breaking patterns presented in this section.

\section{ Summary and    Discussion \label{sec:conclusion}      }

In this paper we have constructed magnetic  monopoles
 arising in spontaneously broken gauge theories  with several symmetry breaking patterns $G \to H$. 
Our work is based on the results of our predecessors \cite{Lb}-\cite{LWY},   and should   be
regarded  as a continuation of their efforts.  Some overlap and repetition of the results discussed here with the earlier ones  \cite{Lb}-\cite{CJH} is
inevitable.
  Nevertheless,  we believe that the present  work constitutes a qualitatively new  contribution to this chain of developments.

Most importantly,  the conditions under which these nonabelian monopoles  survive quantum effects have been clarified, as is discussed in Sections~\ref{NAMonopoles} 
and \ref{sec:quantum}.  In particular there is   now strong   evidence from ${\cal N}=2$  theories that    these monopoles become  light due to  quantum effects and
emerge as the dominant  low-energy degrees of freedom  in certain vacua;  it was pointed out that   the presence of  an appropriate number of massless flavors
in the underlying theory is crucial for this to happen.   As illustrated in $SU(N)$
examples in Section~\ref{NAMonopoles}  and discussed in a more general context in Section~\ref{sec:quantum},    a consistent, quantum theory of nonabelian monopoles 
can be constructed by
 embedding the system   in  ${\cal N} =2$ supersymmetric    theories coupled to an appropriate set of   hypermultiplets.

Although  we  considered in this paper   unconfined monopoles mainly,   a    rather surprising hint about   the nature of nonabelian monopoles 
comes from considering the systems in which the      ``unbroken" group $H$    is further  broken by some other VEVs   at a much smaller mass scale.  
The  nonabelian  monopoles whose gauge bundles
 are described by a transition function in $\pi_1(H) $ are confined by the nonabelian 
{\it vortices} whose squark profile wraps the same class in
$\pi_1(H)$, in the Higgs phase of the $H$ theory. In Ref.~\cite{ABEKY} it
is demonstrated that the vortices of the low-energy system are described by a continuous
 family of solutions transforming under an
irreducible  representation of the dual group ${\tilde H}$.   The monopole-vortex   flux matching
 argument presented in \cite{monovort} and  
proven explicitly for the case 
\begin{equation}    SU(N+1)    \,\,\,{\stackrel {v_1}{\longrightarrow}}     \,\,\, {SU(N) \times U(1) \o  {\mathbb Z}_N  }  
  \,\,\,{\stackrel {v_2}{\longrightarrow}}     \,\,0,
\qquad  v_1 \gg v_2,
\end{equation}
implies that   the monopoles corresponding to all classes in $\pi_1(H)$     are  indeed   confined.   
{\it The existence of a continuous degenerate  family of the vortices  $\pi_1(H)$ implies a   corresponding,  continuous 
family of their sources: the monopoles.}     Most interestingly,  the existence of the quantum mechanical  vortex  zero modes also requires nontrivial
flavors:      the dual group   itself     involves   the original   flavor  group \cite{ABEKY,monovort}.

Three complementary approaches:  (i) topological arguments in which monopole field configurations are classified 
by homotopy groups (Section~\ref{sec:homotopy}); (ii)   general formulae \`a  la GNO-Weinberg, making
use of  the root system structures of the groups  (\ref{sec:General}),  
and   (iii) the explicit construction of BPS solutions  (Section~\ref{sec:Explicit}), have
been used in the present  paper.   Each of these approaches has some advantage over the others.   The (topological)
stability of the monopoles as well as the 
$U(1)$ charge  as compared with the minimum Dirac quantum  are    best explained in the first approach.  On the other 
hand,  the   multiplet structure
 of the minimal  monopoles, and the reason why they  form an irreducible multiplet of   the dual group 
${\tilde H}$,  is  best  understood     in the second approach and easily seen in the third. 
 Finally the
explicit construction (iii)  provides a check of  the whole program   and furthermore is the simplest  way (except for
 MQCD, which yields the same answers) to compute the monopole masses and representations. 
The calculation of the $U(1)$ magnetic charge is explained in \ref{sec:General}.

The results for the quantum numbers of the minimal nonabelian monopoles  in  various  cases are   summarized in
 Table~\ref{monopoles}.  
Note that in some cases   the minimal nonabelian monopoles do not belong to the fundamental
 representation of the unbroken (dual) 
group.   

  \begin{table}[h] 
\begin{center}
{\footnotesize 
\begin{tabular}{| c | c | c  | c |  c |   }
\hline {\bf G} & {\bf H} & {\bf Dual Group} & {\bf Irrep} & {\bf $U(1)$}
\\  \hline  $SU(N+1)$ & $SU(N) \times U(1)/ {\mathbb Z}_N$  &
$SU(N)\times U(1)$  &  $ {\underline N}$    &      $1/N$
\\  \hline  *   ~ $SU(N+1)$    &  $SU(r) \times U(1)^{N-r}/ {\mathbb Z}_r$       &
$ SU(r) \times U(1)^{N-r+1}$   & $ {\underline r}  $     &      $1/r$     
\\  \hline    $USp(2N+2) $ &   $USp(2N) \times
U(1)  $      & $SO(2N+1)\times U(1)$  & $ {\underline {2N+1}}$     & $1 $
\\  \hline  *   ~ $USp(2N+2)$     &  $SU(r) \times U(1)^{N-r}/ {\mathbb Z}_r$       &
$ SU(r) \times U(1)^{N-r+1}$   & $ {\underline r}  $     &      $1/r$     
\\  \hline  $USp(2N) $ & $  SU(N) \times   U(1)/ {\mathbb Z}_N$    
     & $ SU(N)\times U(1)$ &  $ {\underline  N} $   &      $1/N$      
\\  \hline $SO(2N+3) $ &  $ SO(2N+1) \times U(1)$  
        &  $USp(2N)\times U(1)$ &   $ {\underline {2N}} $  &   $1  $
\\  \hline  $SO(2N+2)$   &   $ SO(2N) \times U(1)$
       &  $SO(2N)\times U(1)$  &  $ {\underline {2N}}$  &    $1$  
\\  \hline $SO(2N)$   &   $ SU(N) \times   U(1)/ {\mathbb Z}_N$      &
  $SU(N)\times U(1)$  & $ {\underline {N(N-1)/  2} }$     &      $2/N$      
\\  \hline  *   ~ $SO(2N)$    &  $SU(r) \times U(1)^{N-r+1}/ {\mathbb Z}_r$       &
$ SU(r) \times U(1)^{N-r+1}$   & $ {\underline r}  $     &      $1/r$     
\\  \hline  $SO(2N+1)$   &    $  SU(N)\times   U(1) / {\mathbb Z}_N$  &
 $SU(N)\times U(1)$  & $ {\underline {N(N+1) / 2} }$        &
 $2/N $      
 \\  \hline  * ~ $SO(2N+1)$   &    $SU(r) \times U(1)^{N-r+1}/ {\mathbb Z}_r$      &
 $ SU(r) \times U(1)^{N-r+1}$   & $ {\underline r}  $     &      $1/r$ 
\\  \hline $SU(N+M)$ & $SU(N)\times SU(M) \times U(1) / {\mathbb Z}_k$ &
 $SU(N)\times SU(M) \times U(1)$ & $({\underline
 N},\underline{\overline M})$ & $1/k$
\\  \hline $SO(2N+2M)$ & $SO(2N)\times U(M)$ & $SO(2N)\times U(M)$ &
$(\underline{2N}, \underline{M})$ & $1/M$
\\  \hline $SO(2N+2M+1)$ & $SO(2N+1)\times U(M)$ & $USp(2N)\times U(M)$ &
$(\underline{2N}, \underline{M})$ & $1/M$
\\  \hline $USp(2N+2M)$ & $USp(2N) \times U(M)$ & $SO(2N+1) \times U(M)$ &
$(2N+1, M)$ & $1/M$
\\ \hline
\end{tabular}
}
\caption{\small  Stable nonabelian magnetic monopoles of minimum
    mass arising from the breaking $G \to H$ and     their
    charges.  The $U(1)$ magnetic charge is given in the unit of the
    minimum Dirac quantum, $ 1/2 \, e_0$, where $e_0$ is the minimum
    electric $U(1)$ charge in the system.  $k$ in the $SU(N+M)$   case    is the least common
    multiple of $N$ and $M$.  The cases with  nonmaximal nonabelian factors (*), $r <N $,    are  qualitatively different from the  
    case with $r=N$  in that  monopoles in the fundamental as well as in the  second-rank tensor representation of $SU(r)$ appear: 
 only the monopoles
in the fundamental representation  of $SU(r)$ are indicated.   Note that  these     do not exist for 
  $r=N$ in the case of   $G=SO(N)$, as most easily  seen from the explicit construction  of  Sec.~\ref{sec:sonexp1}  and 
Sec.~\ref{sec:sonexp2}.      }
\label{monopoles}            
\end{center}

 \end{table}


The pattern of the monopole representations appearing in various cases  
found here can be understood based on the Montonen-Olive duality of
${\cal N}=4$ models.  Because of the self-duality, the representation
of the monopoles under the dual group must be the same as that of the
massive vector bosons with respect to the original, electric group.
For instance, when $SO(2N) \to U(N)$, the massive vector bosons are in
the anti-symmetric tensor of $U(N)$, and so are indeed monopoles.
When $USp(2N+2) \to USp(2N) \times U(1)$, the dual theory breaks from
$SO(2N+3) \to SO(2N+1) \times SO(2)$, and the massive vector bosons
are in the vector representation.  Indeed, we find monopoles in the
vector representation of $SO(2N+1)$.

A more non-trivial example is
the breaking $SO(2N+1) \to   U(N)$, where the electric theory has
monopoles in the rank-two symmetric tensor representation of $U(N)$,
while the magnetic $USp(2N)\rightarrow U(N)$ theory has massive
vectors in the same representation.  The electric $SO(2N+1)$ theory
has massive vector bosons in the fundamental and anti-fundamental
representations of $U(N)$, as well as in rank-two anti-symmetric
tensor representations that are twice as heavy.  On the other hand,
the magnetic $USp(2N)$ theory has monopoles in the fundamental,
anti-fundamental representation with the minimum charge, as well as in
rank-two anti-symmetric tensor with twice as large charge and hence
twice as large mass (they are not in Table~\ref{monopoles} because
they can decay into monopoles in the fundamental representation; see
Section~\ref{sec:USpU}).  This agreement is also manifest in MQCD,
where the representations are determined by the same orientifold
projections, employing O3-planes for ${\cal N}=4$ and O4-planes for
${\cal N}=2$.

Finally, the consistency with  Olive-Montonen duality   can be checked also   in the cases  with two nonabelian factors  in $H$. 
For instance,   in the system  $SU(N+M) \to  SU(N)\times SU(M) \times U(1) / {\mathbb Z}_k$ both  the massive gauge bosons of the dual  theory  
and
the monopoles of the electric theory  are in the     $({\underline
 N},\underline{\overline M})$  representation.   Analogously for the case,  $SO(2N+2M) \to    SO(2N)\times U(M)$.  In the 
case    $ SO(2N+2M+1)  \to SO(2N+1) \times U(M)$, the dual theory   breaks as   $USp(2N+2M)  \to  USp(2N) \times U(M)$,
 generating the massive  gauge bosons in the $({\underline  {2N}}, {\underline {M}})$  representation,  consistently with our description in  terms of the
monopoles in the  electric  theory.    Finally, the monopoles in  the $({\underline  {2N+1}}, {\underline {M}})$       found in the system   $USp(2N+2M)  \to  
USp(2N)  \times U(M)$, has  an equivalent,    dual description in terms of massive gauge bosons of   the dual theory   
$ SO(2N+2M+1)  \to SO(2N+1) \times U(M)$.

Another, independent  consistency check of   our general  results can be made by  using   
 the (Lie algebra) isomorphism between the groups  $USp(4)$ and $ SO(5)$. 
Consider the $USp(4)$ theory with  two different symmetry breaking patterns  
\begin{equation}   \bra \phi  \ket_1
=      \left( \begin{array} {cc|cc}      0  & 0   &    &
  \\   0   & v    & &  \\ 
\hline
 & & 0 & 0   
 \\   & & 0 & -v  
 \end{array} \right), \qquad  USp(4)  \to USp(2) \times U(1), 
\label{breaking1}   \end{equation}
\begin{equation}    \bra \phi  \ket_2
=      \left( \begin{array} {cc|cc}      v  & 0   &    &
  \\   0   & v    & &  \\ 
\hline
 & & -v  & 0   
 \\   & & 0  & -v  
 \end{array} \right), \qquad   USp(4)  \to U(2).    
\end{equation}
In the first system, we find (see  Table~\ref{monopoles}, Section~\ref{sec:Explicit})    that the
 minimal monopoles are in the vector (${\underline 3}$) of the
$SO(3)$ group,  which is dual to $USp(2)$, while  in the second system the minimal monopoles 
are in the fundamental representation  (${\underline 2}$)  of the dual $SU(2)$ group, but there
are  also 
 monopoles in the antisymmetric representation   (${\underline 1}$ in this case) with twice the minimal mass.  

In the $SO(5)$ theory, again one has two inequivalent ways to break the gauge symmetry, 
\begin{equation}   \bra \phi  \ket_1
=      \left( \begin{array} {cc|cc|c}      0  & v   &    & &  
  \\   -v    & 0    & &  &   \\ 
\hline
 & &  0 &  v  &     
 \\   & &  -v  & 0  &   \\     
\hline     
&&&& 0 
 \end{array} \right), \qquad  SO(5)    \to   U(2), 
\end{equation}
 \begin{equation}   \bra \phi  \ket_2
=      \left( \begin{array} {cc|cc|c}      0  & v   &    & &  
  \\   -v    & 0    & &  &   \\ 
\hline
 & &  0  &   0 &      
 \\   & & 0   & 0    &   \\
\hline 
&&&& 0  
 \end{array} \right), \qquad   SO(5) \to  SO(3) \times U(1).  
\label{breaking2}\end{equation}
According to our general results,  the monopoles are  in the symmetric second rank tensor representation 
(${\underline 3}$) of $SU(2)$  in the first system
while  in the second case    they are  fundamentals   (${\underline 2}$)   of   the  dual group  $USp(2)\sim SU(2)$
 of $SO(3)$. In this last case there is also a
non-minimal monopole of mass twice the minimal value. 

 The monopole spectra in the   two theories  thus  agree  completely,  as expected.   Actually, the
correspondence can be shown to be  exact: the first (second) pattern of the symmetry breaking in the 
$USp(4) $ theory corresponds to the second  (first)  type of the adjoint VEV in the $SO(4)$ theory.

\begin{figure}[ht]
\begin{center}
\leavevmode
\epsfxsize 12   cm
\epsffile{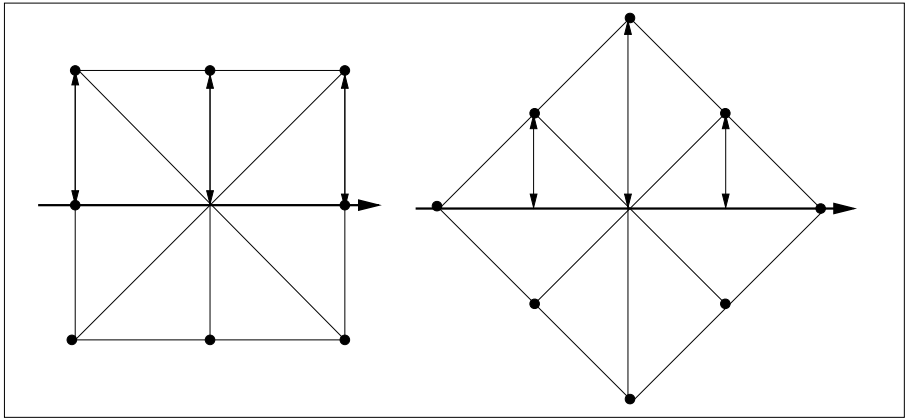}         
\end{center} 
\caption{{\small  Root vectors of  $SO(5)$ (left)  and $USp(4)$ (right).   The root vectors in  the horizontal  directions represent the
 unbroken $SU(2)$ group 
in the case of the first breaking pattern (Eq.(\ref{breaking1})) for $SO(5)$ and the second breaking pattern (Eq.(\ref{breaking2}))  for $USp(4)$.      
The masses of the monopoles are proportional to the heights of the (broken) root vectors. }  }
\label{so5sp4}   
\end{figure}

The whole discussion can actually be  understood very easily  from   the general formula discussed in Appendix A,  as the root vector system
of    the $USp(4) $ group  corresponds simply to a $45$ degree rotation of that of   $SO(5)$ in the bases used above as seen in Figure \ref{so5sp4}. 
The mass spectrum may be read off of this figure as well.

An analogous check, using the isomorphism between  $SO(6)$ and $SU(4)$ which are spontaneously broken respectively to 
$SO(4) \times U(1)$ and   $SU(2) \times  SU(2) \times U(1)$,   also  yields consistent results:    the monopoles are in the ${\underline 4}= ({\underline
2},{\underline 2}  )$  of
$SO(4) \sim  SU(2) \times  SU(2) $.  Another
possibility is to consider the breaking  patterns $SO(6) \to SU(3) \times U(1) $  and  $SU(4) \to SU(3) \times U(1). $  Again,  we learn from
Table~\ref{monopoles}    that the minimal monopoles belong to a triplet (or an antitriplet)  of the unbroken $SU(3)$ in both theories, consistently.

\section* {Acknowledgement} 

We thank N. Dorey, A. Hanany, H. Hansson, K. Higashijima, A. Ritz and
D. Tong for useful discussions and comments.  The work of HM was
supported by the Institute for Advanced Study, funds for Natural
Sciences, as well as in part by the DOE under contracts
DE-AC03-76SF00098 and in part by NSF grant PHY-0098840.  We thank also Sander Bais for 
useful communications.

\appendix

\section{General  Formulae } \label{sec:General}  

In this appendix we review some general formulae \cite{EW,GNO}.
Nonabelian monopoles appear in  a system with  the gauge symmetry breaking 
\begin{equation}     G   \,\,\,{\stackrel {\langle \phi \rangle
     \ne 0} {\longrightarrow}}     \,\,\, H  
\label {general}   \end{equation}
with a nontrivial  $  \pi_2(G/H)$ and $H$ nonabelian.  

The normalization of the generators can be chosen \cite{GNO} so that
the metric of the root vector space is\footnote{In the Cartan basis
  the Lie algebra of the group $G$ takes the form
  \begin{equation}   [H_i, H_k]=0, \qquad (i,k=1,2,\ldots, r);  
    \qquad   [H_i,  E_{\alpha}] = \alpha_i \, E_{\alpha}; \qquad
    [E_{\alpha}, E_{-\alpha}]= \alpha^i  \, H_i; 
  \end{equation}
  \begin{equation}   [E_{\alpha}, E_{\beta} ]=  N_{\alpha \beta}\,
    E_{\alpha + \beta} \qquad (\alpha+ \beta \ne 0).  
  \end{equation}
  $\alpha_i =(\alpha_1, \alpha_2,\ldots)$ are the root vectors.  }
\begin{equation}    g_{ij} =  \sum_{roots} \alpha_i \alpha_j  = \delta_{ij}.   \label{rootnormal} \end{equation}
The  Higgs field vacuum expectation value  (VEV)    is  taken to be of the form 
\begin{equation}     \phi_0  =  {\bf h} \cdot  {\bf H},  
\end{equation}
where  $ {\bf h} = (h_1, \ldots, h_{{\small{\textup{rank}(G)}}})$ 
 is a constant vector representing 
the VEV.        The root vectors  orthogonal to    $ {\bf h}$  belong to the unbroken  subgroup 
$H$.

The monopole solutions are constructed from various  $SU(2)$  subgroups of $G$ that do not commute with $H$,   
\begin{equation}     S_1=  { 1\o \sqrt{ 2 { \bf \alpha}^2}  }  (  E_{{ \bf \alpha}}   +   E_{-{ \bf \alpha}}     ); \qquad  
 S_2=  - { i \o \sqrt{ 2 {\bf \alpha}^2  }    }(  E_{{ \bf \alpha}}   -     E_{-{ \bf \alpha}} ); \qquad  
S_3=   {\bf \alpha}^{*} \cdot  {\bf H }, 
\label{su2g}\end{equation}
where ${\bf  \alpha}$    is a root vector  associated with a pair of    {\it broken}    generators    $ E_{\pm{ \bf \alpha}}$.    
   $\alpha^{*}$ 
is a dual root vector    defined by 
\begin{equation}   \alpha^{*} \equiv  { \alpha \o \alpha \cdot \alpha}.  
\end{equation}
 The symmetry breaking  (\ref{general})  induces the Higgs mechanism in such an $SU(2)$  subgroup, 
$    SU(2)  \to   U(1).
$
By  embedding the known   't Hooft-Polyakov monopole \cite{TP,PS}     lying in  this subgroup and 
 adding a constant term to $\phi$   so that it behaves  correctly  asymptotically,  
one easily constructs a  solution of the 
equation of motion \cite{EW,BK}:  
\begin{equation}   A_i({\bf r})  =  A_i^a({\bf r},  {\bf h} \cdot {\bf \alpha}) \, S_a;  \qquad \phi({\bf r}) =   
  \chi^a({\bf r},  {\bf h} \cdot {\bf
\alpha})
\, S_a   +  [ \,  {\bf h}   -   ({\bf h}
\cdot {\bf \alpha}) \,    {
\bf
\alpha}^{*}  ]  
\cdot {\bf H},  
\label{NAmonop}\end{equation}
where 
\begin{equation}    A_i^a({\bf r}) =  \epsilon_{aij}  { r^j \o r^2}  A(r); \qquad   \chi^a({\bf r}) =  { r^a \o r} \chi(r), \qquad    \chi(\infty)=
  {\bf h} \cdot {\bf \alpha}
\end{equation}
is  the standard   't Hooft-Polyakov-BPS     solution.  Note that $\phi({\bf r}=(0,0,\infty) ) = \phi_0.$  

The mass of a BPS monopole is then  given by 
\begin{equation}      M=\int d{\bf S} \, \cdot    \Tr  \, \phi \,  {\bf B}, \qquad   {\bf B}=  {    r_i  ({\bf S}\cdot {\bf r})  \o r^4}.  \label{mfield}\end{equation}
This    can be computed  by going to the gauge in which 
\begin{equation}  {\bf B}=  {  {\bf r}    S_3    \o r^3}  = {  {\bf r}      \o r^3} \,  {\bf \alpha}^{*} \cdot  {\bf H}, \label{gaugeb}    \end{equation}
to be  
\begin{equation}    M= { 4 \pi   h_i   \alpha^*_j   \o  g  }     \,   \Tr  \, {H_i\, H_j  }.
\end{equation}
For instance, the mass of the minimal monopole of $SU(N+1) \to SU(N)\times U(1)$  
 can be found easily by using Eqs.(\ref{SUNGEN})-(\ref{formula2})
\begin{equation}     M=  { 2 \pi  \, v \, (N+1)  \o g}.
\end{equation}
For the cases $SO(N+2) \to SO(N) \times U(1)$ and $USp(2N+2) \to USp(2N) \times U(1)$, where  
$     \Tr  {H_i\, H_j  } =   C \,   \delta_{ij}, 
$  one finds  
\begin{equation}    M= { 4 \pi   \,  C \,  {\bf h} \cdot  \alpha^*   \o  g  } =    \frac{4 \,\pi \, v}{g  }, 
\end{equation}
while   for  $SO(2N) \to SU(N) \times U(1), $    $SO(2N+1) \to SU(N) \times U(1), $   and $USp(2N) \to SU(N) \times U(1)$, 
the mass is 
\begin{equation}    M= { 8 \pi   \,  C \,  {\bf h} \cdot  \alpha^*   \o  g  } =    \frac{  8  \, \pi \, v}{g  }. \label{masssosu} 
\end{equation}

In order to get the $U(1)$ magnetic charge\footnote{In this
  calculation it is necessary to use the generators normalized as $\Tr
  \, T^{(a)} \, T^{(b)} = {1 \o 2} \delta_{ab},$ such that ${\bf B}=
  {\bf B}^{(0)} \, T^{(0)} + \ldots.  $} (the last column of
Table~\ref{monopoles}), we first divide by an appropriate
normalization factor in the mass formula Eq.(\ref{mfield})
 \begin{equation}      F_m =\int d{\bf S} \, \cdot    {\Tr  \, \phi \,  {\bf B}  \o  N_{\phi}}= \int d{\bf S} \, \cdot  {\bf  B}^{(0)}, \qquad   {\bf B}=  {    r_i  ({\bf S}\cdot {\bf r})  \o
r^4}, 
\label{mflux}\end{equation}
as was done in  Eq.(\ref{Mflux}).  The result,  which is equal to   ${ 4\pi   g_m}$ by definition, gives the magnetic charge.    The latter  must then be expressed
as a function of the  minimum $U(1)$ electric charge   present in the given theory,   which can be easily   found from the normalized  (such that    $\Tr \, T^{(a)}
\, T^{(a)} = {1
\o 2}$)    form of the relevant    $U(1)$  generator.  

   For example, in the case of the symmetry breaking, $SO(2N) \to  U(N)$,   the adjoint VEV is of the form,  $\phi=  \sqrt{4N} \, v \, T^{(0)},$  where $ T^{(0)}$ is
a $2N \times 2N$  block-diagonal  matrix with $N$    nonzero  submatrices $ {i \o  \sqrt{4N} } \pmatrix {0 & 1\cr -1 & 0}.$
Dividing the mass  (\ref{masssosu}) by $\sqrt{N} \, v$  and identifying the flux with $4 \pi g_m$  one gets  $g_m=  { 2 \o  \sqrt{N} \, g}$.   Finally,  in
terms of the minimum  electric charge of the theory  $e_0=   { g \o  \sqrt {4N}}$    ( which follows from the normalized form of $T^{(0)}$ above)   one finds 
\begin{equation}   g_m=   { 2 \o  \sqrt{N} \, g} =    { 2 \o N}   \cdot { 1 \o 2 \, e_0}.  
\end{equation} 
The calculation is similar in other cases.

The asymptotic gauge field can be written as 
\begin{equation}   F_{ij} =  \epsilon_{ijk} B_k = 
\epsilon_{ijk}  { r_k 
\o     r^3}  ({ \beta} \cdot  {\bf H}),    \qquad   { \beta} = {\alpha^*}       \end{equation}
in an appropriate gauge (Eq.(\ref{mfield})).  The Goddard-Nuyts-Olive quantization condition \cite{GNO} 
\begin{equation}  2 \, {\beta \cdot \alpha} \in  { \bf Z} 
\end{equation}
then reduces to the well-known theorem that  for two root vectors  $\alpha_1,\, \alpha_2$    of  any group,
\begin{equation}  { 2 \,  (\alpha_1  \cdot  \alpha_2) \o (\alpha_1  \cdot  \alpha_1) } 
\end{equation}
is an integer.

\section {Root vectors and  weight vectors    \label{sec:Roots}}

\subsection { $A_N=  SU(N+1)$}

It is sometimes convenient to have the root vectors and weight vectors of the Lie algebra $SU(N+1 )$ as vectors in
an $(N+1)$-dimensional space rather than an $N$-dimensional one.  The root vectors are then simply 
\begin{equation}
  (\cdots,   \pm 1, \cdots, \mp 1, \cdots).
\end{equation}
($\cdots$  stand for zero elements)   which all  lie on the plane 
\begin{equation}  x_1+x_2 + \ldots + x_{N+1}=0, \label{sunplane} 
\end{equation}
while the weight vectors are projections in this plane of the orthogonal vectors
\begin{equation}
  \vec{\mu} = (\cdots, \pm1, \cdots)\end{equation}
 where the dots represent zero elements. 

In order to use the general formulas of Weinberg and Goddard-Olive-Nuyts we normalize these vectors so that 
the diagonal  (Cartan)  generators may be written  
\begin{equation}
{\bf{H_i} }  = \diag \,  ( w_{1}^{i}, \,  w_{2}^{i}  \ldots,  w_{N}^{i}, \, w_{N+1}^{i}  \, ),   
 \qquad         i=1,2...N
\label{SUNGEN}  \end{equation} 
where  
   $w_{k}$   represents the $k$-th  weight vector of the fundamental representation 
of $SU(N+1)$, satisfying 
\begin{equation}  {\bf w}_k \cdot    {\bf w}_l =  -  { 1 \o 2 (N+1)^2};  \quad (k \ne l); \qquad   {\bf w}_k \cdot  {\bf w}_k   =   {N \o 2  (N+1)^2},  \qquad
k,l=1,2,\ldots, N+1;
\label{weightsun}\end{equation}
and   $ \sum_{k=1}^{N+1}  {\bf w}_k=0.$  
They are vectors lying  in an $N$-dimensional  space (\ref{sunplane}):  in the coordinates of the
$N+1$-dimensional space,   
\begin{equation}     {\bf w}_i=  { 1\o \sqrt{2 (N+1)^3}}  (-1, \ldots, -1, N,-1,-1,\ldots).   \end{equation}
The root vectors are simply
\begin{equation}   \alpha =  {\bf w}_i -  {\bf w}_j  ={ 1\o \sqrt{2(N+1)} }   (\cdots, \pm 1, \cdots, \mp 1, \cdots) \label{rootssuN}
\end{equation}
with  the norm  \begin{equation}  \alpha \cdot   \alpha = { 1 \o N+1}. \label{normroot}  \end{equation}   
Note that for $i \ne j$
\begin{equation}  \Tr \, (H_i \,  H_j) =  w_1^i w_1^j +    \ldots + w_{N+1}^i w_{N+1}^j =  { -2 N + N-1 \o {2 (N+1)^3}} =  - { 1 \o  2 (N+1)^2},
\label{formula1}\end{equation}
while 
\begin{equation}  \Tr \, (H_i \,  H_i) = {N^2 + N\o  {2(N+1)^3}} =   { N \o 2 (N+1)^2 }.\label{formula2}\end{equation}

The adjoint VEV  causing the symmetry breaking $SU(N+1) \to  SU(N) \times U(1)$ is of the form, 
\begin{equation}  \phi =   {\bf h} \cdot {\bf H},   \qquad  {\bf h}=   v \sqrt{2 (N+1)^3}   \,  (0,0,\ldots,1). 
\end{equation}

\subsection { $B_N  =   SO(2N+1)$}
The  $N$   generators in the Cartan subalgebra  of the Lie algebra $SO(2N+1)$  can be taken to be 
{\small   \begin{equation} 
   H_i =    \pmatrix{ -i  w_1^i {\mathbb J} &  & &  &  \cr
   & - i  w_2^i {\mathbb J} &  & \cr  & & \ddots  &  &    \cr
    &  & & -i  w_N^i  {\mathbb J}  &  \cr  
   &&&&  0   },  \qquad    {\mathbb J}= \pmatrix{ &  1 \cr -1 &} 
\label{songene2} \end{equation} 
}where  ${\bf w}_k$ ($k=1,2, \ldots, N$)  are the weight vectors of the fundamental representation, 
which are     vectors   in  an $N$-dimensional Euclidean space
\begin{equation}  {\bf w}_k \cdot    {\bf w}_l =0;  \quad k \ne l; \qquad     {\bf w}_k \cdot  {\bf
w}_k   =   {1  \o 2(2N-1)}:  \end{equation}
 they form a complete set of orthogonal  vectors. 
The root vectors of  $SO(2N+1)$  group  are   $\alpha=  \{\pm {\bf w}_i, \,\,   \pm {\bf w}_i \pm   {\bf w}_j \}$;
their duals are:
\begin{equation} \alpha^*=   \pm 2(2N-1) \, {\bf w}_i,\qquad   (2N-1) [ \,\pm {\bf w}_i \pm   {\bf w}_j ].
\end{equation}
The diagonal generators satisfy  
\begin{equation}  \Tr \, H_i \, H_j=   {1\o 2N-1}   \,   \delta_{ij}. 
\label{trhihj}\end{equation}
In the system with symmetry breaking  $SO(2N+1) \to SO(2N-1) \times U(1)$  the adjoint scalar VEV is
\begin{equation}  \phi =   {\bf h}\cdot {\bf H},   \qquad  {\bf h}=  i v \sqrt{2(2N-1)} \,  (0,0,\ldots,1). 
\end{equation}

\subsection{$ C_N   =    USp(2N)$}

The  $N$     generators in the Cartan subalgebra  of $USp(2N)$  are the following         $2N\times 2N$ matrices,   
\begin{equation}  {\bf H }_i=     \pmatrix{ {\bf{B_i} }   &  {\bf{ 0 }}  \cr 
  {\bf{ 0 } }   &  - {\bf{B_i}^{t}  }  },     \quad  i=1,2,\ldots,  N, 
\end{equation}
where 
\begin{equation}
{\bf{B_i} }  =
\left( \begin{array}{ccccc}
w_{1}^{i} & & & & \\   
 & w_{2}^{i} & & & \\
 & 0 & \ddots & 0 & \\
 & & & w_{N-1}^{i} & \\  
 & & & & w_{N}^{i}
\end{array} \right) ,   \qquad         i=1,2...N.
\label{USPGEN}   \end{equation} 
The weight vectors  ${\bf w}_k$ ($k=1,2, \ldots, N$)  form a complete set of orthogonal  vectors
  in  an $N$-dimensional Euclidean space and   satisfy 
\begin{equation}  {\bf w}_k \cdot    {\bf w}_l =0;  \quad k \ne l; \qquad     {\bf w}_k \cdot  {\bf w}_k   =   {1  \o 4(N+1)}.  \label{weighusp}\end{equation}
The root vectors of  $USp(2N)$  group  are   $\alpha=  \, \{  \,\pm  \, 2  \, {\bf w}_i, \,\,   \pm {\bf w}_i \pm   {\bf w}_j \}$.
The diagonal generators satisfy  
\begin{equation}  \Tr \, H_i \, H_j=   {1\o 2 ( N+  1)}   \,   \delta_{ij}. 
\label{trhihj2}\end{equation}
For the breaking  $USp(2N) \to USp(2(N-1)) \times U(1)$  the adjoint scalar VEV is
\begin{equation}  \phi =   {\bf h}\cdot {\bf H},   \qquad  {\bf h}=   v \sqrt{4(N+1)} \,  (0,0,\ldots,1 ). 
\end{equation}

\subsection{$ D_N =  SO(2N)$}
The  $N$   generators in the Cartan subalgebra  of the $SO(2N)$  group  can be chosen to be 
{\small   \begin{equation} 
   H_i =    \pmatrix{ -i  w_1^i  \pmatrix{ &  1 \cr -1 &} &  & &  &  \cr
   & - i  w_2^i\pmatrix{ & 1 \cr -1  &} &  & \cr  & & \ddots  &  &    \cr
    &  & & -i  w_N^i\pmatrix{ & 1 \cr -1 &}   &    },  
\label{songene3} \end{equation}  
}where  ${\bf w}_k$ ($k=1,2, \ldots, N$)  are the weight vectors of the fundamental representation, 
living in  an $N$-dimensional Euclidean space and   satisfying
\begin{equation}  {\bf w}_k \cdot    {\bf w}_l =0;  \quad k \ne l; \qquad     {\bf w}_k \cdot  {\bf w}_k   =   {1  \o 4 (N-1)}: \label{weighso2}\end{equation}
they form a complete set of orthogonal  vectors. 
The root vectors of  $SO(2N )$  are   $\alpha=  \{   \pm {\bf w}_i \pm   {\bf w}_j \}$.
The diagonal generators satisfy  
\begin{equation}  \Tr \, H_i \, H_j=   {1\o 2(N-1) }   \,   \delta_{ij}. 
\label{trhihj3}\end{equation}
In the system with symmetry breaking  $SO(2N) \to SO(2N-2) \times U(1)$  the adjoint scalar VEV takes the form
\begin{equation}  \phi =   {\bf h}\cdot {\bf H},   \qquad  {\bf h}=  i v \sqrt{4(N-1)} \,  (0,0,\ldots,1). 
\end{equation}

\end {document}